\documentclass[conference]{IEEEtran}
\usepackage{cite}
\usepackage{amsmath,amssymb,amsfonts}
\usepackage{algorithmic}
\usepackage{graphicx}
\usepackage{textcomp}
\usepackage{xcolor}

\usepackage{circuitikz}
\usepackage{geometry}
\usepackage{array}
\def\BibTeX{{\rm B\kern-.05em{\sc i\kern-.025em b}\kern-.08em
    T\kern-.1667em\lower.7ex\hbox{E}\kern-.125emX}}

\begin{document}

\title{Analog Implementation of the Softmax Function}

\author{\IEEEauthorblockN{Jacob Sillman}
\IEEEauthorblockA{\textit{ECE Department} \\
\textit{University of California, Davis}\\
Davis, CA \\
jlsillman@ucdavis.edu}
}

\maketitle

\begin{abstract}
An analog implementation of the Softmax activation function is presented. A modular design is proposed, scaling linearly with the number of inputs and outputs. The circuit behaves similarly using both a BJT and NMOS design scheme. Experimental results extracted from a BJT breadboard prototype presents computational accuracy within ±4.2\% margin of error. Simulation data presents accuracy within ±1.3\% margin of error for BJT and ±0.9\% margin of error for NMOS design schemes.
\end{abstract}


\section{Introduction}
As the proliferation and expansion of artificial intelligence (AI) continues to evolve into our daily lives, ethical questions of the training process of AI become relevant. Especially in the case of large generalized models, such as large-language models (LLMs) and diverse convolutional neural networks (CNNs), ethical engineers must consider the rising cost of training AI in society across multiple dimensions. Some of the most challenging dimensions of cost is the power and time it takes to train large AI models.

As AI models expand, and almost become incomprehensible from a human scale, so too does the aggregate power consumption of the hardware tasked with hosting the computations that the model is based on. According to OpenAI, GPT-3 consumed as much as 50 PetaFLOP/s-day of compute during pre-training \cite{gpt}. Furthermore, the energy cost associated with hosting the fully-trained model was measured to be at least 1,100 MWh in the month of January 2023 alone. This trend is expected to increase exponentially in the coming years, and engineers must consider unique processor topologies, dynamic systems and methods of computation to reduce the carbon footprint associated with the growing presence of AI in our society.

\subsection{Analog Compute}

There has been significant research in the field of analog computing as a competitor to conventional digital computing. Digital computing has many immediately apparent benefits over analog computing given cursory examination. These benefits include signal abstraction by using discrete voltage levels, relative ease of translation between technology node sizes and robustness to noise and electromagnetic interference (EMI). By contrast, analog systems utilize continuous voltage signals which are inherently sensitive to EMI. As power consumption of an analog system increases, so does the feature size of the system to dissipate the heat. This makes analog systems characteristically difficult to miniaturize.

However, analog compute still has some key intractable features when compared against digital compute:

\begin{itemize}
    \item \textbf{Unclocked computations:} Analog processors are systems whose output nodes find equilibrium that is inherently defined as the solution to the function of initial conditions set by the input nodes. This means single-stage analog processors can compute the output signal which results from a given input perturbation in a single time-constant ($\tau$), independent of class size or load.
    \item \textbf{Ultra-Low Power Operation:} Analog processors have fewer strict requirements on supply voltage and current specifications.This allows analog processors to operate at a much lower power profile than equivalent digital gate logic. As long as all active components maintain the region of operation required for functional processing, the only lower limit on power consumption is the noise floor of the surrounding environment.
    \item \textbf{High Input-Output (I/O) Resolution:} As is inherent in analog signal processing, operating on continuous-time initial conditions results in continuous-time output solutions. Non-discrete solutions allow the system to reach a higher level of cascaded-stage accuracy than digital systems. Wherein digital systems have a limited bit resolution for inputs and outputs, there is no such barrier for analog processors. However, analog processors must still consider the signal-to-noise ratio (SNR) as a soft limit on resolution.
\end{itemize}

With these benefits in mind, we may now address the role which an analog processor may take in technology like AI that is heavily dominated by digital processing.

\subsection{The Softmax Function: A Popular Activation Function}

The procedure of training a CNN model is very iterative. For a distinct task to be performed by the model, a massive amount of training data must be acquired and pre-processed for the model to be trained on.

In the process of forward training, the neural network takes in the input data and applies a series of linear transformations to it, typically represented by matrix multiplications with weights and biases. The output of each layer is then passed through an activation function, which applies a non-linear transformation to the output of the linear transformation. The activation function is a key component of neural networks because it introduces nonlinearity, allowing the model to learn complex relationships between inputs and outputs. Without an activation function, a neural network would essentially be a series of linear transformations, which is not capable of learning complex patterns in the data.

This brings us to the activation function which is the topic of this paper, the Softmax function. The Softmax activation function is commonly used in models that involve multiclass classification problems. Specifically, the Softmax function is used to convert a vector of raw scores, typically generated by the last layer of a neural network, into a probability distribution over a set of classes. In literature, Softmax is typically used for image classification, as it can provide accurate confidence ratings in classifications made by the model when prompted by an input image. The Softmax function is defined as:

\begin{equation}
\sigma(z)_j=\frac{e^{z_j}}{\sum_{k=1}^{K}e^{z_k}}\label{eq1}
\end{equation}

Where $z$ is the input vector of raw scores generated by the previous layers of the network, $j$ is an arbitrary element of that vector and $K$ is the class size of the input and output vectors. The Softmax function computes the exponential of each element of the input vector and normalizes it by the sum of all the exponentials, resulting in a vector of probabilities that add up to 1. This evidence further reinforces the popularity of this activation function for the computation of probability-based analyses. 

In order to compute the Softmax of an input vector, this equation must be solved independently for each element in the input array such that a complete output vector can be produced. Also, all elements in the output vector are correlated with every element in the input vector. This means that a slight change in just one of the input elements can result in a drastic change in all output elements and each element must be recalculated. For large class sizes, this computational task can be taxing. Especially in the case of large models, which are trained with class sizes in the hundreds of millions, the Softmax function can often be a bottleneck of compute time and thus power consumption\cite{bottleneck}\cite{softmaxhard}. 

Using Big-O notation analysis, we can estimate the computational complexity limitations of the function. For arbitrary class size $K$, computing a single Softmax element must sum over all other elements, resulting in complexity of $O(K)$. It should be noted that research in the field of optimization shows that sorting the input vectors in a hierarchical probabilistic order can improve compute complexity to approximately $O(\log_2K)$\cite{bigo}\cite{bigo2}. 

Given this information, the compute-time and power cost of determining the Softmax function of an input vector is a prime candidate for optimization using an analog processor. Since analog processors do not operate on a clock, the only compute delay is due to circuit devices and their effect on the system time-constant.

\section{Circuit Proof}

The circuit concept in this work is based on the differential pair\cite{elf}\cite{txtbook}. The circuit topology proposed can operate with either NPN or NMOS technology. A proof will be presented for both technologies, and then both topologies will be compared for computational accuracy. As first pass proofs, let us assume:

\begin{itemize}
    \item $\beta \rightarrow \infty$
    \item $V_A \rightarrow \infty$ V
    \item $\lambda \rightarrow 0$ V\textsuperscript{-1}
    \item All identical devices are perfectly matched.
\end{itemize}

\subsection{Bipolar Circuit Proof}

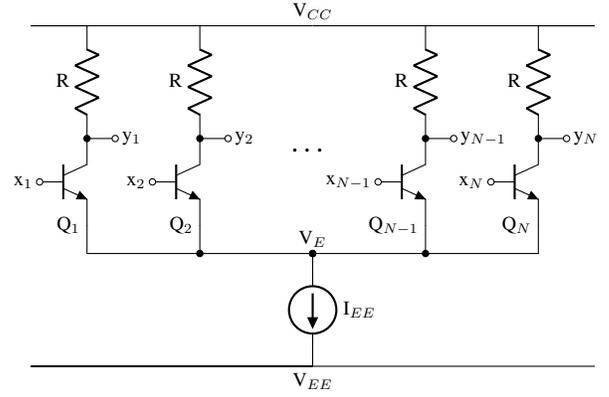
\begin{figure}
    \centering
    \begin{circuitikz}[american, scale = 0.75, transform shape]
         \ctikzset{tripoles/mos style/arrows}
         \ctikzset{transistors/arrow pos=end}
         \ctikzset{legacy transistors text}
         \draw (0,0) node (m1) [npn] {};
         \draw (2,0) node (m2) [npn] {\hspace{1.5cm}\textbf{. . .}};
         \draw (6,0) node (m3) [npn] {};
         \draw (8,0) node (m4) [npn] {};
         \draw (m1.G) to[short,-o] ++(0,0) node [left] {x$_1$};
         \draw (m2.G) to[short,-o] ++(0,0) node [left] {x$_2$};
         \draw (m3.G) to[short,-o] ++(0,0) node [left] {x$_{N-1}$};
         \draw (m4.G) to[short,-o] ++(0,0) node [left] {x$_N$};
         \draw (m1.D) to[short,*-o] ++(0.5,0) node [right] {y$_1$};
         \draw (m2.D) to[short,*-o] ++(0.5,0) node [right] {y$_2$};
         \draw (m3.D) to[short,*-o] ++(0.5,0) node [right] {y$_{N-1}$};
         \draw (m4.D) to[short,*-o] ++(0.5,0) node [right] {y$_N$};
         \draw (m1.S) |- ++(2,-0.5) to[short, *-*] ++(2,0) to[isource, l=I$_{EE}$] ++(0,-2) node (vss) [below] {V$_{EE}$};
         \draw (m2.S) -- ++(0,-0.5);
         \draw (m4.S) |- ++(-2, -0.5) to[short, *-*] ++(-2,0) node[above] {V$_E$};
         \draw (m3.S) -- ++(0,-0.5);
         \draw (m1.S) node [left] {Q$_1$};
         \draw (m2.S) node [left] {Q$_2$};
         \draw (m3.S) node [left] {Q$_{N-1}$};
         \draw (m4.S) node [left] {Q$_N$};
         \draw (m1.D) to[R, l=R] ++(0,2) -- ++(-1,0) -- ++(5,0) node [above] {V$_{CC}$};
         \draw (m2.D) to[R, l=R] ++(0,2);
         \draw (m3.D) to[R, l=R] ++(0,2);
         \draw (m4.D) to[R, l=R] ++(0,2) -- ++(1,0) -- ++(-5,0);
         \draw (vss.north) -- ++(-5,0) -- ++(10,0);
     \end{circuitikz}
    \caption{Modified Differential NPN Network w/ Identical Resistive Loads}
    \label{f1}
\end{figure}

Let us consider an NPN differential circuit. As shown in Figure \ref{f1}, an arbitrary number of bipolar transistors are emitter-coupled to an ideal current source $I_{EE}$. All branches are loaded with a perfectly matched resistive load $R$. Instead of examining this network as a differential network between two output voltages, we evaluate the circuit as a multiple-input-multiple-output (MIMO) system with $N$ independent input and output voltages, which may be expressed as voltage vectors \textit{\textbf{X}} and \textit{\textbf{Y}}.

We assume the input voltages are sufficiently DC-biased to ensure that all transistors are operating in forward-active mode. Since we are assuming $\beta$ is very large, we can neglect base currents ($I_{C,k} = I_{E,k}$). Applying Kirchoff's Current Law (KCL) at the shared emitter node $V_E$,

\begin{equation}
    I_{EE}=\sum_{k=1}^{N}I_{C,k} \label{eq2}
\end{equation}

From \eqref{eq2}, we can expand the expression for collector currents and simplify the equation to get the currents in terms of known node voltages:

\begin{equation}
    I_{EE}=\sum_{k=1}^{N}I_Se^{V_{be,k}/V_T}\label{eq3}
\end{equation}

\begin{equation}
    I_{EE}=\sum_{k=1}^{N}I_Se^{(x_{k}-V_E)/V_T}\label{eq4}
\end{equation}

\begin{equation}
    I_{EE}=\sum_{i=k}^{N}\frac{I_Se^{x_{k}/V_T}}{e^{V_E/V_T}}\label{eq5}
\end{equation}

\begin{equation}
    I_{EE}=\frac{I_S}{e^{V_E/V_T}}\sum_{k=1}^{N}e^{(x_{k})/V_T}=g(V_E)\sum_{k=1}^{N}e^{x_{k}/V_T}\label{eq6}
\end{equation}

It should be noted that emitter node voltage is a variable voltage, and is dependent on the input voltage vector. From here, we can express any branch current as a fraction of the total tail current. The fraction is defined as:

\begin{equation}
    K_i = \frac{I_{C,i}}{I_{EE}}\label{eq7}
\end{equation}

Thus, an arbitrary branch's current can be found:

\begin{equation}
    I_{C,i} = \frac{I_{C,i}}{I_{EE}} I_{EE}\label{eq8}
\end{equation}

\begin{equation}
    I_{C,i}=\frac{e^{x_i/V_T}}{\sum_{k=1}^{N}e^{x_{k}/V_T}} I_{EE}\label{eq9}
\end{equation}

Equation \eqref{eq9} bears a striking resemblance to \eqref{eq1}. The input vector elements take the form of $x_i/V_T$, and the output current is scaled by a factor of $I_{EE}$. We can capture the output current by sampling the voltage across the resistive load $R$. The output voltage would be measured with respect to $V_{CC}$ as opposed to $V_{EE}$. $R$ can be chosen such that $I_{EE}R$ equals 1, and thus the output voltage vector is normalized in the range of one volt. 

While operating in the forward active region, bipolar transistors exhibit an exponential current-voltage relationship, and so the tail current can take any value, and computational accuracy still holds. However, the DC bias point of each of the input voltage elements will shift when the tail current is varied, so care should be taken to choose an application-appropriate value for $I_{EE}$ to ensure that the transistors to do not enter cutoff or saturation.

\subsection{MOS Circuit Proof}

At a glance, one would assume that this circuit would not operate the same way for MOS technologies. This is true for most tail currents, as MOS transistors have a quadratic current-voltage relationship while in saturation. However, for sufficiently small tail currents, this is no longer the case. If we can ensure the transistors operate in subthreshold, we regain the exponential relationship of current\cite{txtbook}: 

\begin{equation}
    I_D=\left(\frac{W}{L}\right)I_te^{(V_{gs}-V_{TH})/nV_T}\left[1-e^{-V_{ds}/V_T}\right]\label{eq10}
\end{equation}

Where $\frac{W}{L}$ is the gate size ratio of the MOS transistor, $I_t$ is the current of the MOS transistor at threshold and $n$ is the subthreshold swing coefficient of the process. The resistive load can be chosen such that the drain-source voltage is sufficiently large ($V_{ds}\gg3V_T$), and we can simplify this current relationship:

\begin{equation}
    I_D=\left(\frac{W}{L}\right)I_te^{(V_{gs}-V_{TH})/nV_T}\label{eq11}
\end{equation}

Now, we can redraw our circuit using MOS devices as shown in Figure 2. A similar mathematical process can used to find the input-output relationship of this MIMO system, starting with KCL at the shared source node. A group constant $C$ is employed for clarity:

\begin{equation}
    C=\left(\frac{W}{L}\right)I_te^{-V_{TH}/nV_T}\label{eq12}
\end{equation}

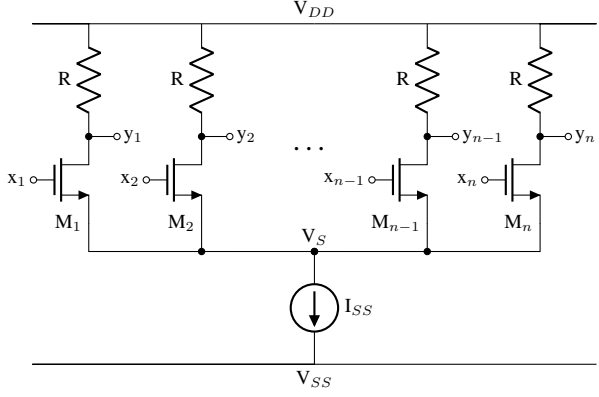
\begin{figure}
    \centering
    \begin{circuitikz}[american, scale=0.75, transform shape]
         \ctikzset{tripoles/mos style/arrows}
         \ctikzset{transistors/arrow pos=end}
         \ctikzset{legacy transistors text}
         \draw (0,0) node (m1) [nmos] {};
         \draw (2,0) node (m2) [nmos] {\hspace{1.5cm}\textbf{. . .}};
         \draw (6,0) node (m3) [nmos] {};
         \draw (8,0) node (m4) [nmos] {};
         \draw (m1.G) to[short,-o] ++(0,0) node [left] {x$_1$};
         \draw (m2.G) to[short,-o] ++(0,0) node [left] {x$_2$};
         \draw (m3.G) to[short,-o] ++(0,0) node [left] {x$_{n-1}$};
         \draw (m4.G) to[short,-o] ++(0,0) node [left] {x$_n$};
         \draw (m1.D) to[short,*-o] ++(0.5,0) node [right] {y$_1$};
         \draw (m2.D) to[short,*-o] ++(0.5,0) node [right] {y$_2$};
         \draw (m3.D) to[short,*-o] ++(0.5,0) node [right] {y$_{n-1}$};
         \draw (m4.D) to[short,*-o] ++(0.5,0) node [right] {y$_n$};
         \draw (m1.S) |- ++(2,-0.5) to[short, *-*] ++(2,0) to[isource, l=I$_{SS}$] ++(0,-2) node (vss) [below] {V$_{SS}$};
         \draw (m2.S) -- ++(0,-0.5);
         \draw (m4.S) |- ++(-2, -0.5) to[short, *-*] ++(-2,0) node[above] {V$_S$};
         \draw (m3.S) -- ++(0,-0.5);
         \draw (m1.S) node [left] {M$_1$};
         \draw (m2.S) node [left] {M$_2$};
         \draw (m3.S) node [left] {M$_{n-1}$};
         \draw (m4.S) node [left] {M$_n$};
         \draw (m1.D) to[R, l=R] ++(0,2) -- ++(-1,0) -- ++(5,0) node [above] {V$_{DD}$};
         \draw (m2.D) to[R, l=R] ++(0,2);
         \draw (m3.D) to[R, l=R] ++(0,2);
         \draw (m4.D) to[R, l=R] ++(0,2) -- ++(1,0) -- ++(-5,0);
         \draw (vss.north) -- ++(-5,0) -- ++(10,0);
     \end{circuitikz}
    \caption{Modified Differential NMOS Network w/ Identical Resistive Loads}
    \label{f2}
\end{figure}

\begin{equation}
    I_{SS}=\sum_{k=1}^{N}I_{D,k}\label{eq13}
\end{equation}

\begin{equation}
    I_{SS}=\sum_{k=1}^{N}Ce^{V_{gs,k}/nV_T}\label{eq14}
\end{equation}

We can see that \eqref{eq14} has the exact same form as \eqref{eq3} from the proof for the bipolar circuit. Therefore the same steps \eqref{eq4} through \eqref{eq8} can be employed to reach a very similar arbitrary branch current solution:

\begin{equation}
    I_{D,i}=\frac{e^{x_i/nV_T}}{\sum_{k=1}^{N}e^{x_{k}/nV_T}}I_{SS}\label{eq15}
\end{equation}

It is clear that the previously defined group constant $C$ is absent from the final solution. It should be noted that this solution does not follow the exact same form as the bipolar solution, due to the incorporation of $n$ in the exponent denominator. The subthreshold swing constant is a function of several technology parameters\cite{subswingparams}. 

In contrast to the bipolar circuit, this topology only functions with vanishingly small currents. The designer must ensure that the current being drawn by the tail current source does not push any one of the signal transistors into saturation mode in order to maintain the exponential current-voltage relationship. Currents in the range of 180 nA to 300 nA were tested in simulation.

As with all signal processing topologies in analog circuits, there are multiple competing interests at play in every design. This too is the case with this analog processor, and the range of considerations vary from basic design constraints such as die area and channel length modulation (CLM) to more complicated and nuanced subjects such as noise analysis and device mismatch. In the next section we will cover aspects of a realizable circuit that an integrated circuit designer must consider when adopting this design. The circuit diagram in Figure 2 is the foundational basis of the design, and it will be developed over multiple iterations as more topics are taken into account to maintain both practical design and computational accuracy.

\section{Basic Design Considerations}

The design of an analog integrated circuit is a complex process that requires careful consideration of several factors to ensure the circuit's proper functionality. This claim is compounded when considering the design of an analog processor, which must maintain an extremely high degree of computational accuracy under a wide variety of input vector conditions. These basic design considerations play a crucial role in ensuring the circuit's stability, performance, and reliability. In this section, we discuss the importance of basic design considerations in the proposed analog processor and how they can impact the circuit's behavior. We highlight the key considerations that must be taken into account during the design process, such as current-source realization, supply voltage margins, CLM and die area and their effects on the circuit's performance. We also explore some design techniques that can be used to address these considerations and optimize the circuit's performance.

\subsection{The Tail Current Source}

The tail current source, labeled as $I_{SS}$ in Figure \ref{f2}, is a critical element in the design. It is very important that the tail current source supplies nearly the exact same magnitude of current over a relatively wide range of voltages at the node $V_S$.

Multiple current mirror topologies were explored for this element. Originally we opted for a Widlar current mirror topology due to its acuity for drawing small and stable currents. Using reasonable biasing parameters for the input voltages, we can expect the shared source node voltage to swing anywhere in the range of 180 mV to 410 mV between different input vectors. At the vanishingly small current magnitude of hundreds of nanoamperes, the systematic gain error (SGE) due to channel-length modulation on the mirror-side of the Wildar current mirror was intolerable -- on the order of hundreds of nanoamperes. In practice, this would essentially cause the tail current source to provide a current that is partially dependent on the source node voltage. This amount of current instability could cause as much as 30\% intrinsic computational error in the design, due to the collapse of the KCL equality in \eqref{eq13}. Due to this discrepancy, we had to choose a more costly current mirror design in order to improve computational accuracy.

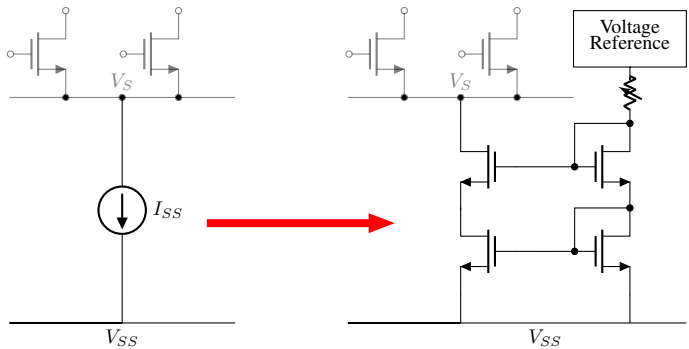
\begin{figure}
    \centering
    \begin{circuitikz}[american, scale = 0.75, transform shape]
         \ctikzset{tripoles/mos style/arrows}
         \ctikzset{transistors/arrow pos=end}
         \draw[opacity = 0.5] (2,0) node (m2) [nmos] {};    
         \draw[opacity = 0.5] (4,0) node (m3) [nmos] {};    
         \draw[opacity = 0.5] (m2.C) to[short, -o] ++(0,0) {};
         \draw[opacity = 0.5] (m2.G) to[short, -o] ++(0,0) {};
         \draw[opacity = 0.5] (m3.C) to[short, -o] ++(0,0) {};
         \draw[opacity = 0.5] (m3.G) to[short, -o] ++(0,0) {};
         \draw[opacity = 0.5] (m2.E) to[short, *-*] ++(1,0) node[above, name = s] {$V_S$};
         \draw[opacity = 0.5] (s.south) to[short, -*] (m3.E) {};
         \draw[opacity = 0.5] (m2.E) -- ++(-1,0) {};
         \draw[opacity = 0.5] (m3.E) -- ++(1,0) {};
         \draw (s.south) to[isource, l={$I_{SS}$}] ++(0,-4) node[below, name=ss]{$V_{SS}$};
         \draw (ss.north) -- ++(-2,0) -- ++(4,0) {};

        \draw[red, line width = 3pt] (4.5,-3) to[short] ++(3,0) node[trarrow, color=red] {};

        \draw[opacity = 0.5] (8,0) node (m2) [nmos] {};    
         \draw[opacity = 0.5] (10,0) node (m3) [nmos] {};    
         \draw[opacity = 0.5] (m2.C) to[short, -o] ++(0,0) {};
         \draw[opacity = 0.5] (m2.G) to[short, -o] ++(0,0) {};
         \draw[opacity = 0.5] (m3.C) to[short, -o] ++(0,0) {};
         \draw[opacity = 0.5] (m3.G) to[short, -o] ++(0,0) {};
         \draw[opacity = 0.5] (m2.E) to[short, *-*] ++(1,0) node[above, name = s] {$V_S$};
         \draw[opacity = 0.5] (s.south) to[short, -*] (m3.E) {};
         \draw[opacity = 0.5] (m2.E) -- ++(-1,0) {};
         \draw[opacity = 0.5] (m3.E) -- ++(1,0) {};
         \draw (9,-2) node (m5) [nmos, xscale=-1] {};
         \draw (9,-3.5) node (m6) [nmos, xscale=-1] {};
         \draw (12,-2) node (m7) [nmos] {};
         \draw (12,-3.5) node (m8) [nmos] {};
         \draw (m7.C) -| (m7.G) {};
         \draw (m8.C) -| (m8.G) {};
         \draw (m5.G) to[short, -*] (m7.G) {};
         \draw (m6.G) to[short, -*] (m8.G) {};
         \draw (m7.C) to[short, *-] ++(0,0) {};
         \draw (m8.C) to[short, *-] ++(0,0) {};
         \draw (m7.D) to[vR, scale = 0.5, xshift=12cm, yshift=-0.6cm] ++(0,0.9) -- ++(0,0.1) -| ++(-1,1) -| ++(2,-1) -- ++(-1,0) node[above, yshift=0.4cm, name=Voltage]{Voltage};
         \draw (Voltage) node[below] {Reference};
         \draw (s.south) -- (m5.D) {};
         \draw (m6.S) -- ++(0,-0.5) node[below, name=ss2] {\hspace{3cm}$V_{SS}$};
         \draw (ss2.north) -- ++(-2,0) -- ++(6,0) {};
         \draw (m8.S) -- ++(0,-0.5) {};
         
     \end{circuitikz}
    \caption{Current Source Realization Utilizing A Cascode Current Mirror}
    \label{f3}
\end{figure}

In the case of this extremely low-power analog processor, SGE can make or break the computational accuracy of this design. With this tenet in mind, we opted for a cascode current mirror. Cascode current mirrors can provide a large range of different current levels with no SGE in theory. Some intrinsic SGE can accrue due to mismatch between transistors in the design, but if the mirror is well matched it can provide a nearly completely stable output current, regardless of shared source node voltage. Of course, this benefit comes at the cost of a higher minimum input voltage at the reference side of the mirror. To provide both a stable and sufficiently large input voltage at the reference side, an apt designer can incorporate a voltage reference of their choosing as well as a method of trimming the reference voltage to tune the design to the desired current magnitude. Simulations of this processor maintain a minimum supply voltage in the range of 600 mV, or approximately 1.5 times the threshold voltage of the process node employed. This claim is in contrast to conventional above-threshold design, which would infer that at least 2 thresholds would be required to ensure both rows of transistors are saturated \cite{txtbook}. However, due to ultra-low current operation, this entire circuit block is operating in subthreshold, so only a fraction of the threshold voltage is required for biasing.

The substitution of the ideal current source by the cascode current mirror is illustrated in Figure \ref{f3}.

\subsection{Channel Length Modulation and Early Effect}

In the case of both the bipolar and NMOS circuit, care must taken when determining how to implement the load resistance for each current branch. If the load is too large, then the amount of voltage swing at the output node will be great enough to cause significant channel length modulation and thus computational inaccuracy.

Therefore, intelligent design decisions must be made to ensure that the loads are both physically small and easy to match, since a full-fledged processor would likely have thousands of branches competing for die area. Depending on the technology node employed, the effect of channel length modulation may vary, so a designer should consult their process design kit (PDK) and characterize their device models before making final decisions on tolerable voltage swing. For the simulations recorded in this work, an output voltage swing of 1 mV is implemented for each branch, meaning that the maximum voltage drop across a branch load is 1 mV. As readers may expect, this decision has both noise and matching implications, both of which are accounted for in the advanced design considerations section. For now, we will assume ideal noise and matching conditions. To translate the output voltage to a more appropriate magnitude, the voltage drop across the load would be sampled through the inverting terminal of a positive-rail tolerant operational amplifier, with the non-inverting terminal connected to $V_{DD}$. For example, if the system specifications required an output voltage range between ground and 1 V, an amplification stage of approximately 60 dB would be employed. 

\begin{figure}
    \centering
    \begin{circuitikz}[american, scale = 0.7, transform shape]
         \ctikzset{tripoles/mos style/arrows}
         \tikzset{mux/.style={muxdemux, muxdemux def={Lh=6, Rh=3, NL=0, NB=0, NR=1}}}
         \draw (1.5,0) node[left] {$\left[ \begin{matrix}x_1 \cr \vdots \cr x_N\end{matrix} \right]$};
         \draw (1.5,0) to[multiwire] (2.5,0) -- (3,0);
         \draw (1.25,-2) node (clk) [oscillator] {};
         \draw (clk.s) node[below] {\footnotesize Clock};
         \draw[dashed] (2.5,3) -- (2.5,-3) -- (12.5,-3) -- (12.5,3) -- (2.5,3);
         \draw (3,0) to[amp, >,boxed, l={\footnotesize Processing Block}, a={\footnotesize Class Size N}, t={$\sigma$}] (4.5,0) to[multiwire] (5.5,0) to[amp,>, t=N, l={\footnotesize Buffer Array}] (7,0) to[multiwire] (8.15,0) node[inputarrow] {};
         \draw (9,0) node[above] {\footnotesize Output};
         \draw (9,0) node[below] {\footnotesize Select Logic};
         \draw (9,0) node[mux]{};
         \draw (10,0) -- (10.5,0) to[amp,>,t={\footnotesize $A$}, l_={\footnotesize Output Amplifier}] (12,0) -- (13,0) node[inputarrow] {};
         \draw (13,0) node[right] {$y_\phi$};
         \draw (11.25,-0.85) node[below] {\footnotesize Gain=60dB};
         \draw (clk.e) -| (6.25,-0.35) node[inputarrow, rotate = 90] {};
         \draw (6.25, -2) to[short, *-] (9,-2) -- (9,-1.3) node[inputarrow, rotate=90] {};
         \draw (9.25,3.5) -| (11.25,0.35) node[inputarrow, rotate=-90] {};
         \draw[dotted] (9.25,3.5) -- (8,3.5) node[above right] {\footnotesize Ext. Output Format Parameters};
           
     \end{circuitikz}
    \caption{Concept of Serialized Output Block Diagram}
    \label{f4}
\end{figure}
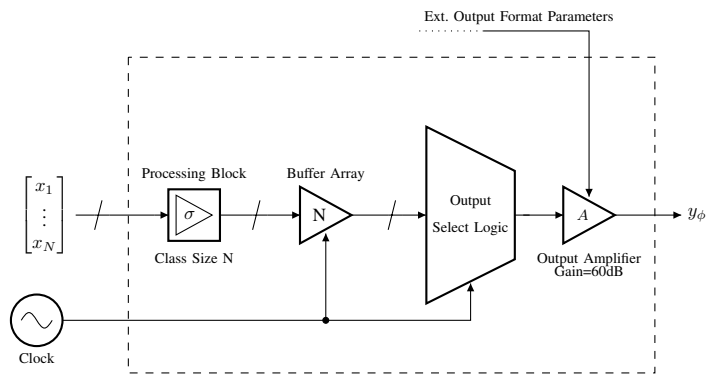

However, implementing a 60-decibel gain stage for each branch would be both impractical and extremely power inefficient. Instead, the output voltage vector would be stored in a capacitive buffer array, which would then be extracted serially using digital select logic. Using this method, only one gain stage would be required to translate the output voltages of every branch. While this solution is more practical, it can greatly reduce the compute speed of the topology, requiring serialization for both inputs and outputs of the processing network. With the introduction of serialization, a clock source must also be introduced to regulate the speed of the digital circuitry. This concept is visualized in Figure \ref{f4}. 

\subsection{Linear-Mode Complementary Loads}

Next, we must consider the form the loads take in each branch, since we want to make sure they take as little area as possible. Simply using resistive polysilicon geometry will likely take too much lateral area on the die, and it would be impractical to have thousands of doped resistive regions on a single die. Moreover, using doped polysilicon resistive elements is often difficult to accurately match and make layout-efficient at the same time.

Taking all of this into account, we would ideally want to use more-easily regulated and compact elements than doped regions, so transistors come to mind. However, we would not want our load transistors to be in saturation, as this would cause uneven gain on each branch and cause the topology to fail. Therefore, using a complementary load that is biased in deep triode would be ideal for this case. Using PMOS loads with the gates all tied to $V_{SS}$, we can guarantee that the transistors are in linear mode. The source-gate voltage will be constant and clearly defined:

\begin{equation}
    V_{SG}=V_{DD}-V_{SS}\label{eq16}
\end{equation}

The current would be controlled by the NMOS signal transistors. This will cause the current through each branch to be much lower than what the PMOS transistors try to draw, which will cause their channels to enter strong inversion and become lightly resistive. We can assume that the souce-drain voltage is very small, so we can make the following approximation for the linear mode drain current:

\begin{equation}
    I_D\approx\left(\frac{W}{L}\right)\mu_pC_{ox}\left[\left(V_{SG}-|V_{TH,P}|\right)V_{SD}\right]\label{eq17}
\end{equation}

Since the biasing of the gate is well-defined, the equivalent resistance of the load can be found:

\begin{equation}
    R_{SD}=\frac{V_{SD}}{I_D}=\frac{1}{\frac{W}{L}\mu_pC_{ox}(V_{DD}-V_{SS}-|V_{TH,P}|)}\label{eq18}
\end{equation}

Therefore, the resistance of the linear mode complementary load is defined by design parameters. PMOS transistors, while bulkier than more conventional NMOS devices, are more compact than integrated resistors and are able to be packed tightly together. Dense layout is likely required for a processor designed to handle input vectors with thousands of elements. 

\begin{figure}
    \centering
    \begin{circuitikz}[american, scale = 0.57, transform shape]
         \ctikzset{tripoles/mos style/arrows}
         \ctikzset{transistors/arrow pos=end}
         \ctikzset{tripoles/pmos style/nocircle}
         \ctikzset{legacy transistors text}
         \ctikzset{bipoles/crossing/size = 0.4}
         \draw (-4,0) node (m1) [nmos] {};
         \draw (-1,0) node (m2) [nmos] {};
         \draw (2,0) node (m3) [nmos] {};
         \draw (5,0) node (m4) [nmos] {};
         
         \draw (-4,1.5) node (m9) [pmos, xscale=-1] {};
         \draw (-1,1.5) node (m10) [pmos, xscale=-1] {};
         \draw (2,1.5) node (m11) [pmos, xscale=-1] {};
         \draw (5,1.5) node (m12) [pmos, xscale=-1] {};
         
         \draw (0.5,-2.5) node(m5) [nmos, xscale=-1] {};
         \draw (0.5,-4) node(m6) [nmos, xscale=-1] {};
         \draw (8,-2.5) node(m7) [nmos] {};
         \draw (8,-4) node(m8) [nmos] {};
         \draw (m1.G) to[short,-o] ++(0,0) node [left] {x$_1$};
         \draw (m2.G) to[short,-o] ++(0,0) node [left] {x$_2$};
         \draw (m3.G) to[short,-o] ++(0,0) node [left] {x$_{n-1}$};
         \draw (m4.G) to[short,-o] ++(0,0) node [left] {x$_{n}$};
         \draw (m1.D) to[short,*-o] ++(0.5,0) node [right] {y$_1$};
         \draw (m2.D) to[short,*-o] ++(0.5,0) node [right] {y$_2$};
         \draw (m3.D) to[short,*-o] ++(0.5,0) node [right] {y$_{n-1}$};
         \draw (m4.D) to[short,*-o] ++(0.5,0) node [right] {y$_{n}$};
         \draw (m1.S) |- ++(3,-0.5) to[short,*-] ++(1.5,0) -- (m5.D) {};
         \draw (m2.S) -- ++(0,-0.5);
         \draw (m4.S) |- ++(-3, -0.5) to[short, *-*] ++(-1.5,0) node[above] {};
         \draw (m3.S) -- ++(0,-0.5);
         \draw (m1.S) node [left] {M$_1$};
         \draw (m2.S) node [left] {M$_2$};
         \draw (m3.S) node [left] {M$_{n-1}$};
         \draw (m4.S) node [left] {M$_{n}$};
         \draw (m5.S) node [left, yshift=0.5cm] {};
         \draw (m6.S) node [left, yshift=0.5cm] {};
         \draw (m7.S) node [right, yshift=0.5cm] {};
         \draw (m8.S) node [right, yshift=0.5cm] {};
         \draw (m9.S) -- ++(0,0.5) -- ++(-1,0) -- ++(7,0) node[above] {V$_{DD}$};
         \draw (m10.S) -- ++(0,0.5) {};
         \draw (m11.S) |- ++(8,0.5) {};
         \draw (m12.S) -- ++(0,0.5) {};
         \draw (m9.G) -- (m10.G) -- (m11.G) -- (m12.G) -| ++(0.2,-3) to[crossing] ++(0,-2) to[crossing] ++(0,-1) -- ++(0,-0.25) {};
         \draw (m7.D) to[short,*-*] ++(0,0) -| (m7.G){};
         \draw (m8.D) to[short,*-*] ++(0,0) -| (m8.G){};
         \draw (m5.G) to[short,-*] (m7.G){};
         \draw (m6.G) to[short,-*] (m8.G){};
         \draw (m7.D) to[vR, scale = 0.5, xshift=8cm, yshift=-1.2cm] ++(0,0.9) -- ++(0,0.1) -| ++(-1,1) -| ++(2,-1) -- ++(-1,0) node[above, yshift=0.4cm, name=Voltage]{Voltage};
         \draw (Voltage) node[below] {Reference};
         \draw (0.5,1) node[above, scale = 2] {\Large\textbf{...}};

         \draw (m6.S) node (vss) [below, xshift=1.5cm] {V$_{SS}$};
         \draw (vss.north) -- ++(-7,0) -- ++(15,0);
     \end{circuitikz}
    \caption{Updated Circuit Diagram w/ Current Mirror and Linear-Mode Load}
    \label{f5}
\end{figure}

We will see in Section IV, specifically when we consider noise performance, that another form of complementary load will be better for noise performance as well as simplifying output vector extraction. Figure \ref{f5} shows an updated topology that takes into account both the updated linear-mode complementary loads and a realized current mirror. The voltage reference circuit is omitted for simplicity, as many voltage reference generation topologies may be adequate for the circuit. In simulation, the voltage reference is replaced with an ideal current source.

\subsection{Die Area and Supply Margin}

Much of the compromises of die area is intertwined with other factors, such as load choice and branch complexity and has been discussed in other sections at length. Ideally, we wish to use as little die area as possible. In this analog processor, each branch operates in parallel, which means that the processor can only compute the Softmax solution for class sizes equal-to or less-than the number of branches fabricated on-chip.

For example, a processor that has one thousand branches can compute a solution of a one thousand element input-vector by utilizing all branches on-chip. If the same processor wishes to compute a class size of only five hundred, it can tie the latter 500 branch inputs to $V_{SS}$ and still compute the Softmax solution with equal accuracy as the prior case.

With this in mind, it makes sense that we should try to minimize the die area contribution of each branch. For each square micron we save per branch, we can afford to include more branches and thus increase the range of class sizes the processor can tolerate. Therefore, all transistors on-chip should utilize minimum channel-length transistors for the process node employed, as well as the smallest number of fingers for each transistor in the branch. This rule of thumb doesn't necessarily apply to blocks located outside the array of branches, such as the current mirror or voltage reference, but care should be taken to minimize the area of all blocks on the die to improve maximum class-size acceptance.

There are some benefits to using subthreshold devices on the issue of supply voltage margin. In subthreshold, device biasing does not require an entire threshold voltage. Depending on the current a transistor drives, it can have many different bias voltages. In this way, the MOS topology operates very similarly to a bipolar circuit. Through the negative feedback introduced by the tail current source, much of the work of intelligently biasing each transistor to receive current proportional to the accurate Softmax solution is provided by the shared source node. Due to this fact, the source node voltage can take a wide variety of different values depending on the input vector applied. As presented by \cite{txtbook}, the minimum voltage required at the source node for the current mirror to operate is given by:

\begin{equation}
    V_{src,min}=V_{TH} + 2V_{OV}\label{eq19}
\end{equation}

Where \eqref{eq19} assumes all transistors operate in the saturation region. However, in subthreshold, the required source voltage changes to:

\begin{equation}
    V_{in,min}\approx2V_{TH,sub}\label{eq20}
\end{equation}

Where $V_{TH,sub}$ is the threshold voltage of each transistor on the reference-side of the mirror \cite{cascode}. Depending on the technology node employed and reference current chosen by the designer, this variable has a definite value. With this in mind, we know the definite minimum voltage of the source node is approximately $2V_{TH,sub}$. This number can be reduced by either reducing the reference current or increasing the transconductance of the mirror by increasing the number of fingers. However, the benefits obtained by increasing these parameters should be balanced with the drawbacks of increased die area and noise performance.

Moving up the diagram, the maximum voltage at any of the output nodes can be as high as the supply voltage. The lowest voltage would be the supply voltage minus the maximum load swing, which we will call $V_{swing}$ for this analysis. As explained previously, we wish to maintain the drain-to-source voltage of the signal transistors to be significantly larger than $3V_T$ to neglect the effect of leakage swing. A reasonable value to choose would be to limit the drain-source voltage to be at least 200 millivolts. Now, we can compute the approximate supply voltage margin:

\begin{equation}
    V_{DD,min} \approx 2V_{TH,sub}+V_{swing}+0.2 V\label{eq21}
\end{equation}

This supply margin is lower than many comparable circuits, since all active transistors are in subthreshold. This supply voltage is small enough to be biased with sub-one volt supplies, which makes it both extremely low power and versatile. For the simulations conducted in this work, a supply voltage of 1.8 volts was employed.

 Skilled electrical engineers know that there are many design choices that must be made that consequently create more problems in another realm of analysis. It is the duty of an intelligent designer to weigh the costs and benefits of each choice in the grand scheme of the system and choose the best option possible. In this section, we covered design challenges that had fairly straightforward solutions with little cost to the overall operation and accuracy of the processor.  In the following section, we will cover more advanced topology design considerations that involve more complex insights into how the processor operates.

 \section{Advanced Design Considerations}

In the last section, much of the discussion was centered on effects that will impact the accuracy of the circuit. This section will cover more effects on accuracy, but more importantly issues that stem from resolution. In digital circuits, resolution is simply how many bits it takes to represent a signal. In the case of analog signal processing, resolution is closely tied to signal-to-noise ratio (SNR). How well one can differentiate between a circuit's solution and noise is integral to having a useful processor. Wherein some analog circuits a designer may be happy with an SNR of 10 to 100, a sensitive and high-accuracy analog processor must have a much more strict requirement on noise suppression. For the function to be solved within reasonable accuracy with the circuit described in this work, we aim for an SNR in the range of 100 to 1000 (40 dB to 60 dB).

\subsection{Input Vector Tolerance}

We live in a world dominated by digital processing. With this fact, if we wish to implement an analog processor, we must convert our input data to analog voltages. With the conversion of digital data to analog voltages, there comes undesired effects. As is typical in digital-to-analog converters (DACs), voltages are generated using mixed-signal circuitry that may carry harmonics and oscillations introduced by clock sources. Even if we assumed that the analog voltage inputs were generated perfectly with no harmonics or noise introduced, there is still the effect of parasitic cross-talk between different parts of the processor that can create harmonics or oscillations with frequencies on the order of many unique time constants that are intrinsic to the circuit. With the understanding that non-ideal input voltage elements are unavoidable in a practical realization of this circuit, we must consider what effect harmonics and oscillations in the input voltages will have on the accuracy and resolution of the processor. For our analysis, we assume that the harmonics and oscillations introduced are significantly lower than the transit frequency ($f_t$) of the process node employed. This simplifies the analysis to only consider how unstable inputs affect the Softmax solution, and puts less emphasis on the effect of input frequency on the poles and input capacitance of the topology.

Since the function being solved in this circuit is non-linear, a given amplitude of noise at the input can relate to different levels of error at the output. In this analysis, we consider two different types of computation -- a ``well-matched" compute and a ``single-dominant" compute.

In a well-matched compute, some or all of the input vector elements to the processor are very similarly biased, meaning the output vector elements share similarly-even portions of the total output. A simple example of a well-matched compute for a class size of $N=4$:

\begin{equation}
    \left[ \begin{matrix}1.00 \cr 1.00 \cr 1.00 \cr 1.00\end{matrix} \right] \xrightarrow{\sigma} \left[ \begin{matrix}0.25 \cr 0.25 \cr 0.25 \cr 0.25\end{matrix} \right]\label{eq22}
\end{equation}

\begin{figure}[htbp]
    \centering
    \includegraphics[scale = 0.75]{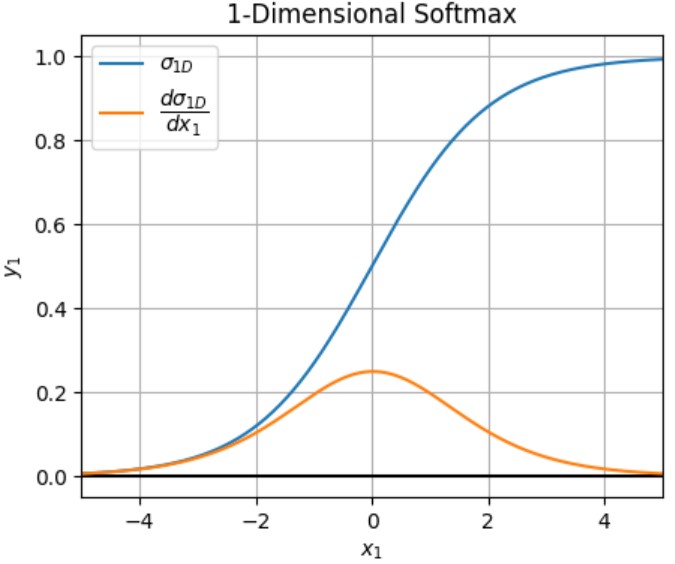}
    \caption{Plot of two functions. In blue, the one-dimensional Softmax function. In orange, the gradient of the one-dimensional Softmax function with respect to the input $x_1$.}
    \label{f6}
\end{figure}

If we apply input-side voltage noise to one of the input elements in this configuration, the noise has a large magnitude effect on the output vector. This is due to the fact that the gradient of the 4-dimensional Softmax function at this point is very steep. We define the set of well-matched computes as input vectors that lie in the region surrounding this sharp gradient point. To illustrate this point, Figure \ref{f6} shows a one-dimensional Softmax function, otherwise known as a sigmoid function. $x_2$ assumes the value of 0, and $x_1$ is swept between -5 and 5. Along with the sigmoid plot, its derivative is also shown, which is the gradient of the Softmax function with respect to $x_1$. We can see that when $x_1\approx x_2$, the slope of the gradient is greatest, which means that small undesired harmonics will have the greatest effect at this point. The radius around this sharp gradient point is somewhat arbitrarily chosen, as it depends heavily on the relative spacing of input vector elements, class size and circuit components. To give relevance to this kind of compute in context, an input vector that would match the description of a well-matched compute would be inferences that a machine learning model would generate early in forward training, when it has somewhat even confidence in all of the possible class elements.

In a single-dominant compute, a singular input vector element is sufficiently larger than all other input elements relative to the average size of all input element magnitudes. This would mean that one of the input vector elements would receive the vast majority share of the output vector, leaving the rest close to zero. Once again, an example of a single-dominant compute for a class size of $N=4$:

\begin{equation}
    \left[ \begin{matrix}1.00 \cr 5.00 \cr 1.00 \cr 1.00\end{matrix} \right] \xrightarrow{\sigma} \left[ \begin{matrix}0.02 \cr 0.94 \cr 0.02 \cr 0.02\end{matrix} \right]\label{eq23}
\end{equation}

Under this configuration, input-side voltage noise applied to any of the vector elements has far less of an effect on the output vector elements. By the same reasoning as before, this point on the 4-dimensional Softmax function has a much shallower gradient. This means that the input vectors have a higher tolerance to input-side harmonics, oscillations and noise in this configuration than in the case of well-matched computes. This set of input vectors is more common in typical neural network operation, specifically when the network more easily converges to a solution with high confidence. Of course, well-matched computes still arise regularly when the neural network is experiencing difficulty with a particular inference task.

Very little can be done on the side of the hardware designer to ensure that single-dominant computes occur more often than well-matched computes. However, designers can introduce input and output node filtering to minimize the amplitude of undesired harmonics and noise when they inevitably arise.

\subsection{Integrated Components: Noise Analysis}

Noise analysis is an important step in the design process for all integrated and discrete circuits. However, in the case of ultra-low power circuits, noise analysis is especially important as the noise signals that are being generated by circuit components can have comparable size to the magnitude of the signals being operated upon for the circuit to function. Therefore, any reduction we can make to noise figure is critical to supporting ultra-low power operation. 

In circuit analysis, noise is often expressed in terms of power spectral density (PSD) instead of voltages or currents. This is because noise is a stochastic process that has a random variation over time and frequency, making it difficult to analyze directly in the time domain. By using the PSD, which is the Fourier transform of the autocorrelation function of the noise \cite{stanfordnoise} , the power distribution of the noise across different frequencies can be analyzed. The PSD is measured in units of power per frequency, typically in V\textsuperscript{2}/Hz or A\textsuperscript{2}/Hz. By analyzing the PSD, it is possible to determine the frequency range in which the noise is most prominent and design circuits that can minimize or filter out noise in that range. Additionally, using PSD allows us to calculate the total noise power in a system, which is important for determining the noise figure of an amplifier or other circuits.

When considering components that generate noise, engineers typically employ the Johnson-Nyquist noise model \cite{mitnoise}. The Johnson noise model can predict noise that originates from multiple independent sources. The most widely-studied noise sources that are incorporated in the model are as follows:

\begin{itemize}
    \item \textbf{Thermal Noise}: This is typically the largest noise-contributor in integrated circuits. Thermal noise is caused by the thermal agitation of charge carriers, which generates small fluctuations in the voltage and current of a conductor. These fluctuations are random and follow a White Gaussian Noise (WGN) distribution, meaning that they have equal power across all frequencies. The magnitude of thermal noise is proportional to the square root of the temperature of the conductor, and the bandwidth of the system over which it is measured. Thermal noise can be expressed as either a Norton and Th\'evenin equivalent circuit. For transistors operating in saturation, thermal noise can be expressed as a spectral current source:

    \begin{equation}
        \overline{i_n^2}=\frac{4k_BTg_m}{3}\Delta f\label{eq24}
    \end{equation}

    Where $k_B$ is Boltzmann's constant, $g_m$ is the transconductance of the transistor at its bias point and $\Delta f$ is the bandwidth with which the processor is being sampled. Note that the thermal noise scales directly with temperature and gain.

    \item \textbf{Flicker Noise}: Unlike thermal noise, which is caused by the random motion of electrons at non-zero temperature, flicker noise has no clear physical origin and is not well understood. However, it is believed to be caused by a combination of factors, including the random trapping and release of charge carriers in impurities or defects in the material, and the fluctuation of surface charges in the device. The main characteristic of flicker noise is that it has a spectral density that is inversely proportional to frequency, meaning that the noise power increases as the frequency decreases. Flicker noise is typically most prominent at low frequencies and due to this fact is often also called ``1/f noise." Flicker noise is present in transistors regardless of operation region, and is expressed as a spectral current source:

    \begin{equation}
        \overline{i_n^2}=\frac{K_1I_D}{f}\Delta f\label{eq25}
    \end{equation}

    Where $K_1$ is the flicker constant for the device technology.

    \item \textbf{Shot Noise}: Shot noise is a type of noise that arises from the discrete nature of charge carriers in electronic systems. As electrons flow through a potential barrier in integrated devices, they move in a random, stochastic manner due to the probabilistic nature of quantum mechanics. This randomness leads to fluctuations in the current or voltage of the device, which can be modeled as a current or voltage source with a spectral density that is proportional to the square root of the current or voltage. This noise is known as shot noise because it arises from the ``shots" of individual electrons moving through the device. The magnitude of shot noise depends on the average current or voltage of the device, as well as the bandwidth of the measurement. Shot noise is typically measured with respect to the DC gate current, but it is also measured in the drain during subthreshold operation \cite{stanfordnoise}. This is due to the fact that the channel in weak inversion acts as a potential barrier similar to a bipolar device. In the same way the gate oxide acts as a thin and tall potential barrier, the subthreshold channel acts as a short and long potential barrier. In subthreshold, shot noise spectral current is defined as:

    \begin{equation}
        \overline{i_n^2}=2qI_D\Delta f\label{eq26}
    \end{equation}

    Where $q$ is the electron charge. Typically in the gate, this noise acts as a much smaller component compared to other noise sources. In subthreshold, shot noise is certainly larger due to a higher drain current magnitude compared to gate current, but still smaller than other noise sources by-and-large.
    
\end{itemize}

Since none of the transistors in the Softmax topology operate in the saturation region, charge carriers do not generate thermal noise in our signal transistors. However, the PMOS loads in our system operate in linear, which still generate thermal noise. While transistors operate in strong inversion, the conducting channel operates as a low-impedance resistor which can be thermally agitated in the same way as a typical polysilicon integrated resistor. The linear-mode thermal noise can be expressed as a voltage source:

\begin{figure}
    \centering
    \begin{circuitikz}[american, scale=0.7, transform shape]
         \ctikzset{tripoles/mos style/arrows}
         \ctikzset{transistors/arrow pos=end}
         \ctikzset{tripoles/pmos style/nocircle}
         \ctikzset{legacy transistors text}
         \draw (3,0) node (m1) [pmos, scale=2]{};
         \draw (m1.G) to[short, -o] ++(0,0) {};
         \draw (m1.D) to[short, -o] ++(0,0) {};
         \draw (m1.S) to[short, -o] ++(0,0) {};
         \draw (3.5,0.6) node[flowarrow, rotate=-90, anchor=west, scale=2]{\rotatebox{90}{$I_D$}};
         \draw (1,-4) to[R, l={$R_{SD}$}, f_={$I_D$}] ++(0,-2) to[short, -*] ++(2,0) to[short, -o] ++(0,-0.5){};
         \draw (3,-4) to[isource, l={$\overline{I_{Th}^2}$}] ++(0,-2) {};
         \draw (5,-4) to[isource, l={$\overline{I_{1/f}^2}$}] ++(0,-2) -- ++(-2,0){};
         \draw (1,-4) to[short, -*] ++(2,0) -- ++(2,0) {};
         \draw (3,-4) to[short, -o] ++(0,0.5){};
         
         \draw (7,0) to[R, l={$R_{SD}$}, f_={$I_D$}] ++(0,-2) {};
         \draw (9,0) to[isource, l={$\overline{I_{1/f}^2}$}] ++(0,-2) {};
         \draw (7,0) to[vsource, l_={$\overline{V_{Th}^2}$}] ++(0,2) to[short, -o] ++(0,0) {};
         \draw (7,0) to[short, *-] ++(2,0){};
         \draw (9,-2) to[short, -*] ++(-2,0) to[short, -o] ++(0,-0.5){};

        \draw[blue, line width = 3pt] (5,-0.5) to[short] ++(1,0) node[trarrow, color=blue] {};
        \draw[green, line width = 3pt] (3,-1.8) to[short] ++(0,-1) node[trarrow, color=green, rotate=-90] {};

     \end{circuitikz}
    \caption{In linear region, thermal noise can be expressed as either an equivalent spectral voltage or spectral current source, as they pertain to the Norton (in green) and Th\'evenin (in blue) circuit models.}
    \label{f7}
\end{figure}
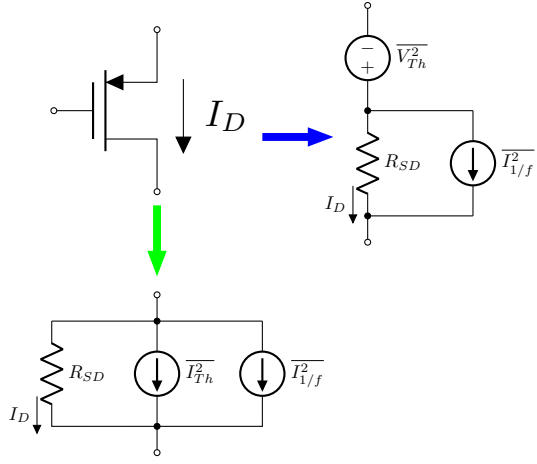

\begin{equation}
    \overline{v_n^2}=4k_BTR_{SD}\Delta f\label{eq27}
\end{equation}

Where $R_{SD}$ is the resistance seen across the conducting channel, which was expressed previously in \eqref{eq18}. Figure \ref{f7} shows two ways to model the thermal noise in a PMOS device while in the ohmic region. In the case of our noise analysis, we can isolate a singular output branch and study the effects of noise sources on the output node. It is easier to treat the thermal noise generated by the PMOS device as a spectral voltage source as it is directly connected to our branch output node. Since the DC current is the same for both the PMOS and NMOS devices in the branch, both elements contribute a similar amount of flicker noise. The flicker constant may vary between the devices, but the overall magnitude is fairly constant.

\begin{figure}
    \centering
    \begin{circuitikz}[american, scale = 0.7, transform shape]
         \ctikzset{tripoles/mos style/arrows}
         \ctikzset{transistors/arrow pos=end}
         \ctikzset{legacy transistors text}
         \draw (-2,0) -- (0,0) node[above]{$V_{DD}$};
         \draw (2,0) -- (0,0) {};
         \draw (0,-2) to[vsource, l={$\overline{V_{Th}^2}$}] ++(0,2) {};
         \draw (0,-2) to[short,*-] ++(-2,0) to[isource, l={$\overline{I_{1/f}^2}$}] ++(0,-2) to[short,-*] ++(2,0) -- ++(0,-0.5) to[short,*-o] ++(0.5,0) node[right]{$V_{out}$};
         \draw (0,-2) to[R,l={$R_{SD}$}] ++(0,-2) to[short,-*] ++(0,-1) to[isource, l={$I_D$}] ++(0,-2) to[short,*-o] ++(0,-0.5) node[below] {$V_S$};
         \draw (0,-5) -- ++(-2,0) to[isource, l={$\overline{I_{1/f}^2}$}] ++(0,-2) -- ++(2,0) {};
         \draw (0,-5) -- ++(2,0) to[isource, l={$\overline{I_{shot}^2}$}] ++(0,-2) -- ++(-2,0) {};

     \end{circuitikz}
    \caption{Circuit diagram of an arbitrary Softmax branch, with transistors replaced with their equivalent noise models.}
    \label{f8}
\end{figure}
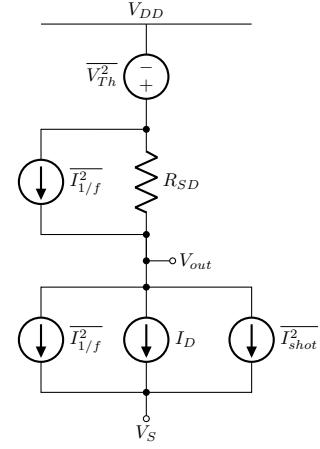

The noise-equivalent models can be replaced in the circuit diagram of an arbitrary branch as shown in Figure \ref{f8}. For the most part, the noise contribution due to flicker noise cannot be avoided. The best a designer can do is try to operate their processor at a higher frequency or use a device technology with a sufficiently low flicker constant. The desired output signal we wish to extract can be given by:

\begin{equation}
    V_{out}=I_DR_{SD}\label{eq28}
\end{equation}

However, by KCL, a resistance-scaled noise term is also generated by the shot noise in the signal transistor:

\begin{equation}
    \overline{v_n^2}=\overline{I_{shot}^2}R_{SD}^2\label{eq29}
\end{equation}

This noise is also combined with the thermal noise from the PMOS load:

\begin{equation}
    \overline{v_n^2}=\overline{V_{Th}^2} \label{eq30}
\end{equation}

As is often the case in noise analysis, it can be helpful to consider the root-mean-square (RMS) of the spectral noise sources, which allows us to more easily consider the noise sources as operable circuit components. The spectral sources can be replaced with direct components by taking the square-root of the magnitude of the spectral sources. Under this condition, we can express the RMS of the output noise voltage:

\begin{equation}
    \overline{V_{noise}} = \overline{I_{shot}}R_{SD} + \overline{V_{Th}}\label{eq31}
\end{equation}

\begin{equation}
    \overline{V_{noise}} = R_{SD}\sqrt{2qI_D\Delta f} + \sqrt{4k_BTR_{SD}\Delta f}\label{eq32}
\end{equation}

Conventionally, to improve SNR, designers may decide to increase the magnitude of the desired signal compared to the noise signal. In the case of our signal branch, there are two ways to do this.

If our goal is to increase the magnitude of our signal while adding as little to the magnitude of the noise as possible, we should tune both the branch current $I_D$ and the load impedance $R_{SD}$, as can be seen in \eqref{eq28}. From \eqref{eq32}, we can easily see that the size of $R_{SD}$ doubly contributes to the gain of the noise. It contributes linearly in the shot noise calculation and in a root-law manner in the thermal noise calculation. Conversely, $I_D$ only contributes in a root-law manner. From this view, we should consider scaling $I_D$ as much as possible. This would scale $V_{out}$ linearly, but only scale $\overline{V_{noise}}$ in a square-root manner, overall improving SNR.

However, there is a limit to how large we can scale the branch current. We want to ensure that the tail current does not exceed our subthreshold condition, as when any of the signal transistors enter saturation, they no longer compute the Softmax solution. As defined in \cite{txtbook}, a MOS transistor enters saturation mode when the drain current surpasses $I_t$, which they express as: 

\begin{equation}
    I_t=qXD_nn_{po}e^{k_2/V_T}\label{eq33}
\end{equation}

Where $k_2$ is a function of a constant of integration $k_1$. Other texts express subthreshold current using other variables and constants, but this work uses the definition outlined in \cite{txtbook} for modelling purposes. However it should be noted that the transition from subthreshold to saturation is not piecewise, and is better understood as a gradient between the two conduction states. Therefore, it is typically better to stay well below $I_t$ to preserve as much computational accuracy as possible.

Additionally, adding more current to the system increases the overall power consumption of the topology, which may not be viable for certain ultra-low power applications.

Using this reasoning, we can only increase $I_D$ so much before we run into computational inaccuracy. Therefore, it is difficult to increase the swing of the output voltage without increasing the load impedance, consequently linearly increasing shot noise. There are other issues with trying to increase $R_{SD}$. For example, trying to create a larger linear impedance will involve needing to increase the length of the transistor. This adds to the total area of the branch, which as discussed previously, can greatly increase die area. Alternatively, a designer may increase the bias voltage at the gate of the PMOS load, which would require developing a biasing circuit that could also increase the total area of the topology and add more complexity. Additionally, adding more swing across the transistor will also cause a greater magnitude of $V_{SD}$, which contests the conditions made in the approximation from \eqref{eq17}. This makes the impedance across the resistor less constant over the range of possible values of $V_{out}$, which skews our computational accuracy.

However, there are issues with increasing the magnitude of the output voltage in general. As was expressed in the basic design considerations, we want to limit the drain node voltage swing as much as possible in order to reduce the variation in drain current due to CLM. By increasing the swing of $V_{out}$, we are also greatly increasing the effect of CLM on our Softmax solution. This makes increasing the output magnitude far less viable as a means of improving SNR.

Another common strategy to improving the SNR of an amplifier is to add capacitance to the load. This adds low-pass filtering to the output, which reduces the effects of high-frequency noise at the output. Designers must be careful when choosing a capacitance value for the output node, because adding capacitance increases the computation time of the topology. The capacitor takes time to charge to different voltage levels, which takes a few time constants to reach equilibrium. The time constant can be reduced by using smaller load impedance, smaller capacitance or higher drain current. This strategy has a similar issue as increasing load impedance, as die area is required to fabricate integrated capacitors, and if the system specification calls for a Softmax processor with a thousand branches, then the area cost of the integrated capacitors would be increased a thousand-fold.

We can consider some approximate numbers for $I_D$ and $R_{SD}$ to get estimates for the magnitude of the noise contributions of each noise source. Using $R_{SD}$ on the order of $10^3$ $\Omega$ and $I_D$ on the order of $10^{-7} A$, we can express the noise contribution of each noise source as a percentage:

\begin{equation}
    \overline{V_{shot}} \text{ contribution} \approx4.21\%\label{eq34}
\end{equation}

\begin{equation}
    \overline{V_{thermal}} \text{ contribution} \approx95.79\%\label{eq35}
\end{equation}

We can recognize that the thermal noise contributes far more to the total noise than shot noise. Therefore, if we can implement a load that only operates in subthreshold, then we can reduce the total noise of our system by a significant degree without needing to increase output swing.

\begin{figure}
    \centering
    \begin{circuitikz}[american, scale=0.8, transform shape]
         \ctikzset{tripoles/mos style/arrows}
         \ctikzset{tripoles/pmos style/nocircle}
         \ctikzset{transistors/arrow pos=end}
         \ctikzset{legacy transistors text}
         \draw (0,-2) node (m1) [pmos, xscale=-1] {};
         \draw (m1.D) to[short,-o, f_={$I_D$}] ++(0,-1) {};
         \draw (m1.S) to[short,-o] ++(0,0) {};
         \draw (m1.G) -| ++(0.5,-1) node[ground] {};

         \draw (4,0) node (ma) [pmos, xscale=-1] {};
         \draw (4,-1.5) node (mb) [pmos, xscale=-1] {};
         \draw (mb.S) node[below left, yshift=0.1cm] {$W_1$};
         \draw (8,0) node (mc) [pmos] {};
         \draw (8,-1.5) node (md) [pmos] {$W_2$};
         \draw (mb.D) to[short,*-o, f_={$I_D$}] ++(0,-1) {};
         \draw (mb.D) -| (mb.G) {};
         \draw (md.G) to[normal open switch] ++(-1.5,0) to[short, -*] (mb.G) {};
         \draw (mc.G) to[normal open switch] ++(-1.5,0) to[short, -*] (ma.G) |- (ma.D) to[short,-*] ++(0,0) {};
         \draw (md.D) to[short,-*, f_={$\frac{W_2}{W_1}I_D$}] ++(0,-1) to[short,-o] ++(0.5,0) node[right] {$V_{out}$};
         \draw (md.D) -- ++(0,-1.5) to[short,*-] ++(0.5,0) to[R, l=$R_{load}$] ++(0,-2) to[short, -*] ++(-0.5,0) node[ground] {};
         \draw (md.D) -- ++(0,-1.5) to[short,*-] ++(-0.5,0) to[C, l_=$C_{load}$] ++(0,-2) -- ++(0.5,0){};
         \draw (ma.S) to[short,-o] ++(0,0) {};
         \draw (mc.S) to[short,-o] ++(0,0) {};
     \end{circuitikz}
    \caption{Left: The original PMOS linear-mode resistive load. Right: Updated shared PMOS cascode current mirror load with digital select MOS switches.}
    \label{f9}
\end{figure}
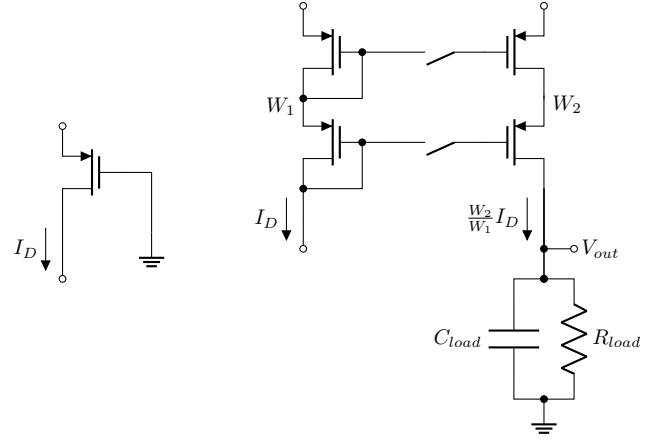

Figure \ref{f9} shows a shared PMOS cascode current mirror load that can be implemented in the topology. We replace the PMOS operating in linear mode with a cascode current mirror that is attached to an output branch that is loaded with an integrated resistor and capacitor. This strategy transfers the accurate Softmax current to a separate branch where it can be used to drive a larger load without needing to worry about large output swing. All branches are connected to this current mirror, but are gated by MOS switches which are controlled by digital select logic. Only one branch is switched at any given time, and this allows us to digitally select which output current we wish to copy to the output branch, and also transitions most of the high-impedance and large-area components to a single non-repeated load branch. This drastically reduces our area and impedance constraints, since we only need to fabricate a single load for every branch to share. It is also a helpful change because it now allows the analog output voltage vector to be measured with respect to ground instead of $V_{DD}$, which can be easier to interpret for other connected systems.

Also, this topology replaces the single linear mode device with the equivalent of four subthreshold-mode devices. As shown in \eqref{eq34} and \eqref{eq35}, this exchange is worth the added complexity when considering the magnitude of noise contributions from transistor devices. However, there is still thermal noise present in this design due to the integrated resistor load. However, this design gives us more control over how we can manipulate the thermal noise. First, by including the capacitor in parallel, we can filter and integrate a lot of the WGN due to both thermal and shot noise. Designers should note that this comes at the cost of compute speed. Additionally, the PMOS transistors from the cascode current mirror can have their widths scaled to add linear gain to the copied branch current, which allows us to tune both $I_D$ and $R_{load}$ in tandem without needing to worry about any loss to computational accuracy. The copied current can be amplified by the current mirror width ratio with almost no consequence, but there is a limit to the size a designer can make the PMOS channel area. If one makes the width too long, the computation speed will be limited by how long it takes to charge up the PMOS gate capacitance. Also, just as the drain current is amplified in the output branch, so too is the noise contribution from the reference branch transistors.

Of course, this solution does not come without its drawbacks. Adding a PMOS cascode current mirror to the top of the circuit, as well as having an NMOS cascode current mirror at the bottom of the circuit makes the full topology five transistors tall. This will significantly increase the supply voltage margins, adding two more $V_{TH,sub}$ to the minimum supply margin. Also, copying the current and potentially multiplying it by an additional current mirror width ratio will introduce increased power consumption. 

This PMOS cascode current mirror load is not the only way to improve SNR. Designers may decide to use one of the other loads proposed in previous iterations of the circuit.

There exists an optimal balance between power consumption, die area, output impedance size, Softmax current amplification, compute speed, computational accuracy and noise figure. An astute designer must account for each of these variables in their design when fabricating a practical implementation of this processor and draw their own conclusions for the best possible allocation that the full system integration calls for.

\subsection{Integrated Components: Mismatch Analysis}

Recall the circuit proof for the NMOS configuration outlined in \eqref{eq10} through \eqref{eq15}. This proof only holds when we assume that there is perfect matching between all signal transistors. In a real full-system implementation, this is never the case.

When modelling mismatch, we must consider each parameter that is susceptible to variation between transistors. In the case of the variables in \eqref{eq11} and \eqref{eq33}, we can expect device mismatch in $X$, $D_n$ and $\frac{W}{L}$ during fabrication. We can combine the variation of these elements in a process-variation group constant $\Delta c$ similar to \eqref{eq12}. We can then re-examine \eqref{eq14} as:

\begin{equation}
    I_{SS}=\sum_{k=1}^N(C+\Delta c_k)e^{V_{gs,k}/nV_T}\label{eq36}
\end{equation}

\begin{equation}
    I_{SS}=\sum_{k=1}^NCe^{V_{gs,k}/nV_T}+\sum_{k=1}^N\Delta c_ke^{V_{gs,k}/nV_T}\label{eq37}
\end{equation}

Now, accounting for mismatch, we can also rewrite \eqref{eq15} and expand:

\begin{equation}
    I_{D,i}=\frac{(C+\Delta c_i)e^{x_i/nV_T}}{\sum_{k=1}^NCe^{x_{k}/nV_T}+\sum_{k=1}^N\Delta c_ke^{x_{k}/nV_T}}\label{eq38}
\end{equation}

Now, unfortunately there is nothing we can do to mathematically simplify this expression. However, if we assume that the device mismatch of each transistor is Gaussian and that the process variation on the die area is relatively uniform, we can distribute the transistors uniformly about the die surface. If we sum over the mismatch of all components, we can roughly approximate:

\begin{equation}
    \sum_{k=1}^N\Delta c_ke^{x_{k}/nV_T} \rightarrow 0\label{eq39}
\end{equation}

Using this rough approximation, we can rewrite \eqref{eq38}:

\begin{equation}
    I_{D,i}=\frac{(1+\frac{\Delta c_i}{C})e^{x_i/nV_T}}{\sum_{k=1}^Ne^{x_{k}/nV_T}}I_{SS}\label{eq40}
\end{equation}

As we can see, under the best case conditions of distributing the signal transistors over uniform process variation and assuming Gaussian mismatch, the branch current accuracy has a linear relationship with device mismatch. For example, with $\pm1\%$ $\Delta c$ group-constant mismatch, we would expect at most $\pm1\%$ computational error for any given output vector element.

Now, with all design considerations accounted for, we can begin looking at simulation and experimentation data recorded for this proposed work.

\section{Simulation Setup}

Verification of this circuit design was conducted in multiple simulation programs. The bipolar circuit was verified and simulated in LTSpice and Cadence PSpice. LTSpice was also used to select circuit components for the experimental breadboard prototype circuit. Most circuit simulations were conducted in Cadence Virtuoso. The circuit was designed using 180-nm predictive technology models from the Predictive Technology Group. Cadence Virtuoso was used to verify design optimization choices at multiple stages throughout the processor design process. ADE-L and Cadence Spectre was used to run transient, DC sweep and noise-analysis simulations to measure circuit accuracy.

\subsection{Measuring Processor Accuracy}

On the topic of circuit accuracy, there are many complications that arise when it comes to measuring the accuracy of a Softmax processor. One way to measure accuracy is to bias the input nodes with preset voltages that we know the correct Softmax solution for and subtract the difference between the expected solution and the extracted solution. However, as expressed in Section IV-A, the magnitude of error can vary between different input vectors. Measuring the true accuracy of an analog processor is an extremely comprehensive process, one too large to tackle for the scope of this work. In this work, we measured accuracy in a simpler way.

We measured output solution accuracy by setting $N-1$ of the processor inputs to a median DC value. Then, we took the last branch and swept the input node through a range of voltages around the DC value while measuring the same branch's output node voltage. On a mathematical basis, we are collapsing the N-dimensional Softmax function down to a single dimension. This makes the input-output relationship resemble a sigmoid function, which we can then compare to an ideal sigmoid curve under identical scale and horizontal shift conditions to measure computational accuracy.

Once again we should note that this accuracy test is in no way comprehensive, and only shows us a tiny slice of the true accuracy of the processor. However, this accuracy test is both easy to implement and fast to test in both simulation and in prototype experimentation.

\section{Simulation Results}

\subsection{Bipolar Simulations}

First, looking at the bipolar circuit simulations, we constructed the same circuit in both LTSpice and PSpice. All bipolar transistors use the ZTX1048A model. This model was chosen due to its large $h_{FE}$ gain and high $V_A$ voltage. Additionally, the ZTX1048A was readily available on Mouser to construct the physical breadboard prototype circuit. In the case of all circuits simulated (both NMOS and NPN), a class size of $N=4$ was chosen. In the case of the bipolar circuit, the loads were simply implemented using resistors. Also, the cascode current mirror was replaced with a simple current mirror.

\begin{figure}
    \centering
    \includegraphics[width=\linewidth]{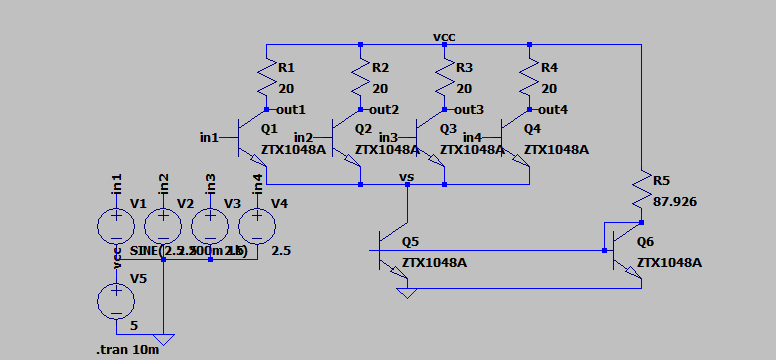}
    \caption{LTSpice Schematic of the Bipolar Processor Topology}
    \label{f10}
\end{figure}

\begin{figure}
    \centering
    \includegraphics[width=\linewidth]{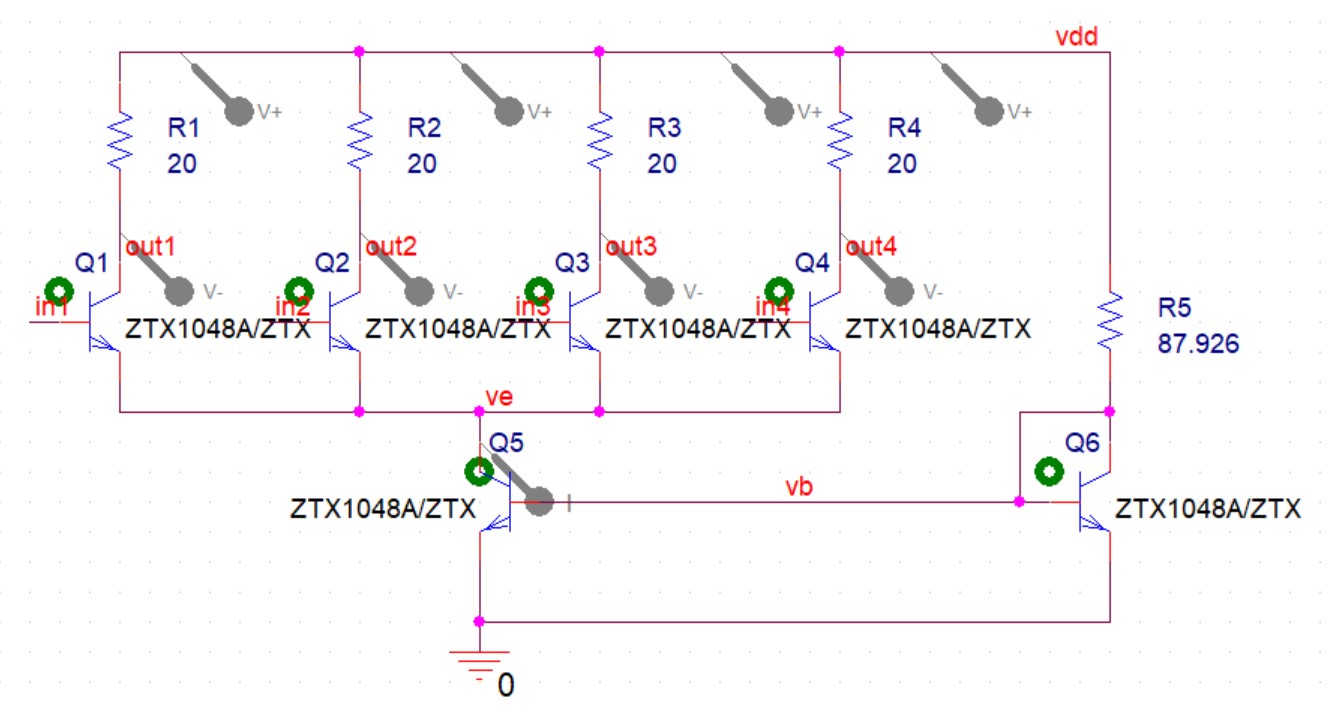}
    \caption{Cadence ORCAD PSpice Schematic of the Bipolar Processor Topology}
    \label{f11}
\end{figure}

Figures \ref{f10} and \ref{f11} show the simulation schematics from LTSpice and PSpice, respectively. Both schematics look identical, as the only variable between the two simulations is the compiler used to run them. A resistor was used in place of a current source in these simulations, as a resistor will be used later on to drive the tail current mirror in the physical prototype board.

\vspace{0.25cm}
\begin{center}
\textbf{Design Parameters for Both Bipolar Processor Circuits}
\vspace{0.1cm}
    \begin{tabular}{|c|c|}
       \hline
       Design Parameter  &  Value\\
       \hline
       \hline
       Load Impedance  &  $20$ $\Omega$\\
       \hline
       Output Voltage Swing  &  1V\\
       \hline
       Tail Current  &  50 mA\\
       \hline
       Supply Voltages  &  $V_{CC}=5V, V_{EE}=0V$\\
       \hline
       Input DC Bias Voltage  &  2.5V\\
       \hline
       Class Size  &  4\\
       \hline
    \end{tabular}
\end{center}

\vspace{0.25cm}

\begin{figure}
    \centering
    \includegraphics[width=\linewidth]{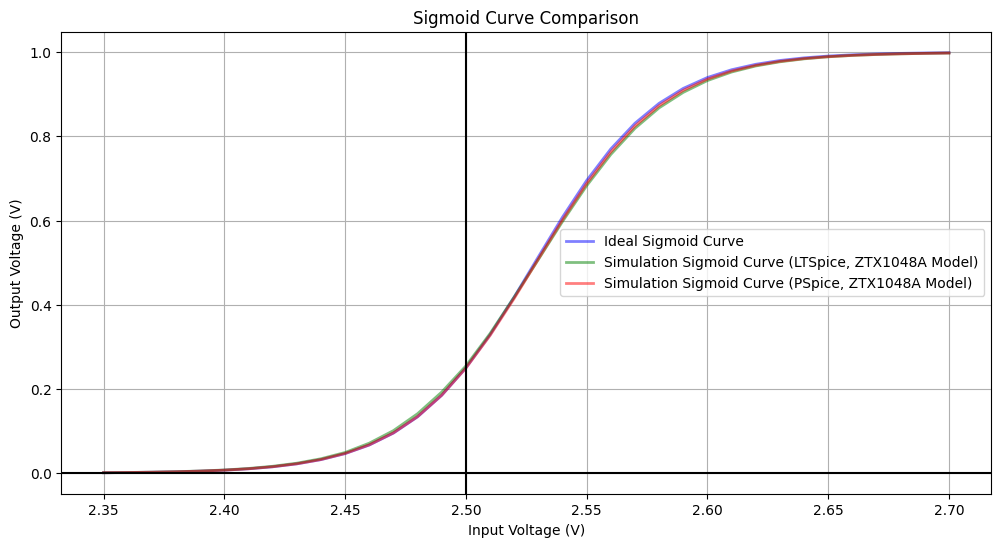}
    \caption{Sigmoid Curve Comparison for LTSpice and PSpice Bipolar Simulations}
    \label{f12}
\end{figure}

\begin{figure}
    \centering
    \includegraphics[width=\linewidth]{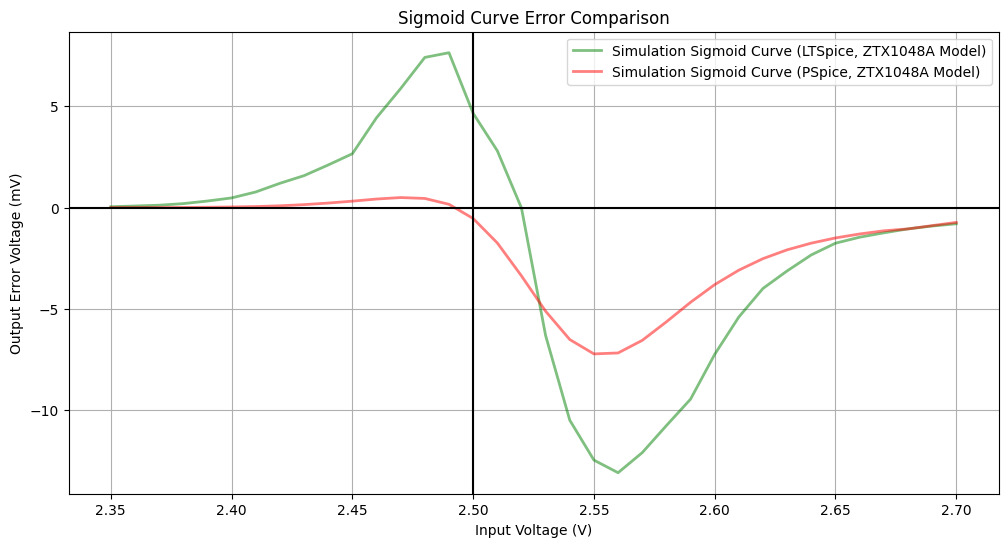}
    \caption{Sigmoid Error Comparison for LTSpice and PSpice Bipolar Simulations}
    \label{f13}
\end{figure}

Figure \ref{f12} shows the extracted sigmoid sweep curves from branch 1 for both simulation profiles. These sigmoid curves are plotted alongside the ideal sigmoid curve in blue. It is very difficult to tell the difference between the three curves because the large signal accuracy is very high. The accompanying lower plot in Figure \ref{f13} shows the inherent error in the circuit relative to the ideal sigmoid curve. As you can see, the error in the bipolar circuit is very small. This is due to the fact that bipolar transistors have an exponential relationship between bias voltage and collector current in forward active mode, so there is no upper current limit to the processor's computational validity. Some of the error is easy to explain. There is a slight negative gradient in the error curves that is likely due to the finite early voltage of the transistor model. As the output voltage swings towards one volt, the magnitude of the collector voltage decreases (because the output voltage is measured with respect to $V_{CC}$) and therefore $V_{CE}$ is reduced. This causes early effect on the transistor to be lessened, and thus draw less parasitic CLM current. The remaining error that is present is likely due to a combination of higher order effects pertaining to channel effects and carrier mobility as well as the non-zero base current flowing from the input nodes. The maximum recorded error in the bipolar circuit implementation does not exceed $\pm1.3\%$.

The bipolar circuits that were simulated did not use low-power parameters, due to using the exact same circuit elements in a lab experiment setting. In order for the output voltage to be distinguishable over noise on an educational lab oscilloscope, a tail current of 50 mA was chosen. This caused the power consumption of this topology to be on the order of 500 mW.

\subsection{NMOS Simulations}

All NMOS simulations were designed and captured using Cadence Virtuoso. Some simulations were recorded using resistive loads for simplicity, and other simulations used the low-noise load configuration. Tail current was set to a range of different values between 180 nA and 300 nA. Inspect the table below for other design parameters:

\vspace{0.25cm}
\begin{center}
\textbf{Design Parameters for NMOS Processor Circuit}
\vspace{0.1cm}
    \begin{tabular}{|c|c|}
       \hline
       Design Parameter  &  Value\\
       \hline
       \hline
       PMOS Casc. Width Ratio  &  1.00\\
       \hline
       Load Resistance  &  3.5 M$\Omega$\\
       \hline
       Load Capacitance  &  50 fF\\
       \hline
       Output Voltage Swing  &  1 V\\
       \hline
       Drain Voltage Swing  &  1 mV\\
       \hline
       Tail Current  &  $180$ nA $\le I_{SS} \le 300$ nA\\
       \hline
       Supply Voltages  &  $V_{DD}=1.8$ V, $V_{EE}=0V$\\
       \hline
       Input DC Bias Voltage  &  600 mV\\
       \hline
       Subthreshold Swing Coef.  &  $n = 1.71$\\
       \hline
       Compute Frequency  &  $250$ kHz\\
       \hline
       Class Size  &  4\\
       \hline
    \end{tabular}
\end{center}

\vspace{0.25cm}

\begin{figure}
    \centering
    \includegraphics[width=\linewidth]{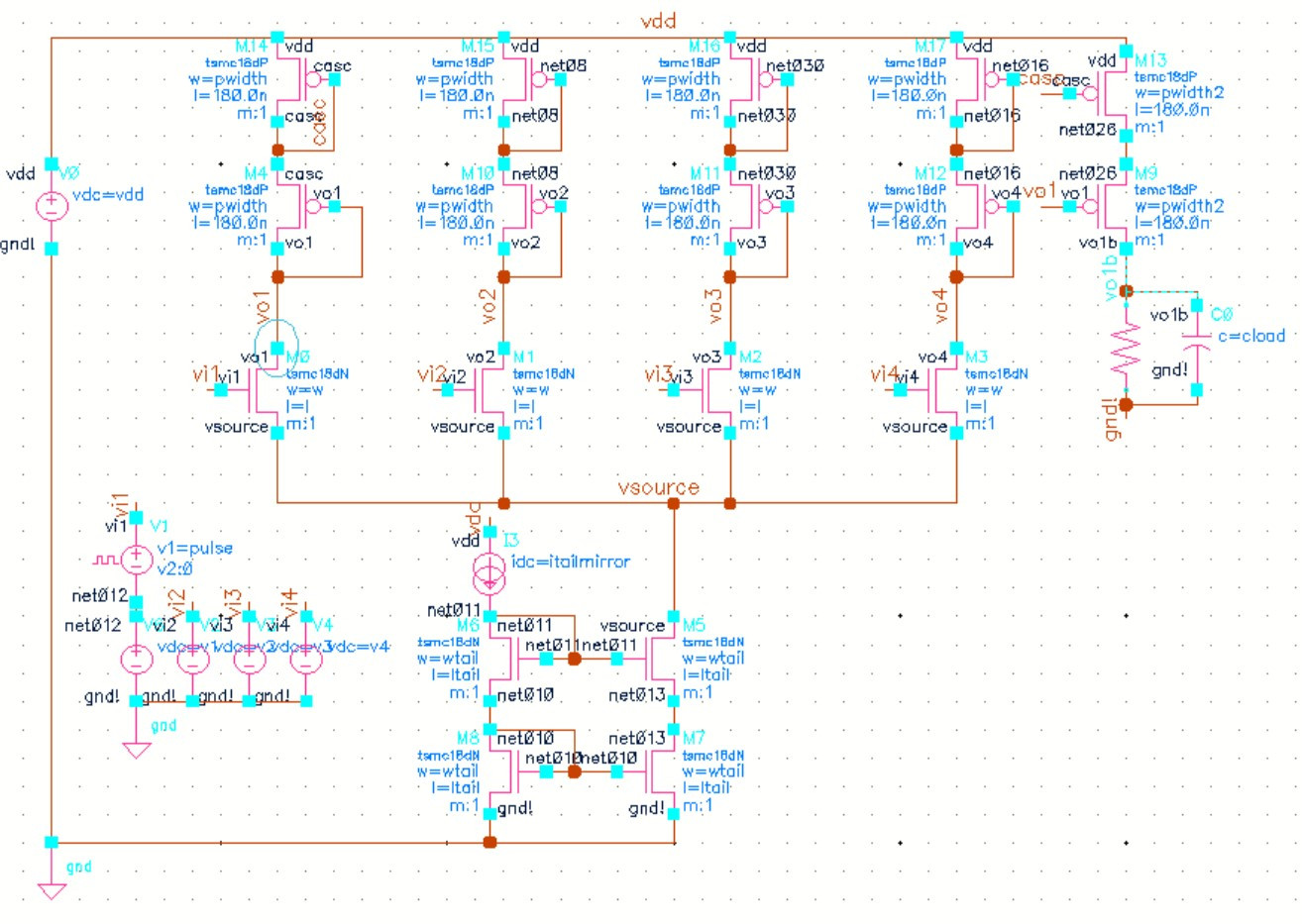}
    \caption{Cadence Virtuoso Schematic of the NMOS Processor Topology}
    \label{f14}
\end{figure}

Figure \ref{f14} shows the Virtuoso schematic diagram for the full topology, including the low-noise load. Many transistor widths and other circuit parameters were replaced with design variables that could be tuned and optimized in the Spectre analog test bench between simulations. As mentioned previously, the reference tail current was generated with an ideal current source for ease of use. The first branch's input voltage was swept between 400 mV and 900 mV, which is wider than the range for the bipolar circuit. This is due to the larger magnitude of the denominator in the Softmax exponent due to the subthreshold swing coefficient.  

\begin{figure}
    \centering
    \includegraphics[width=\linewidth]{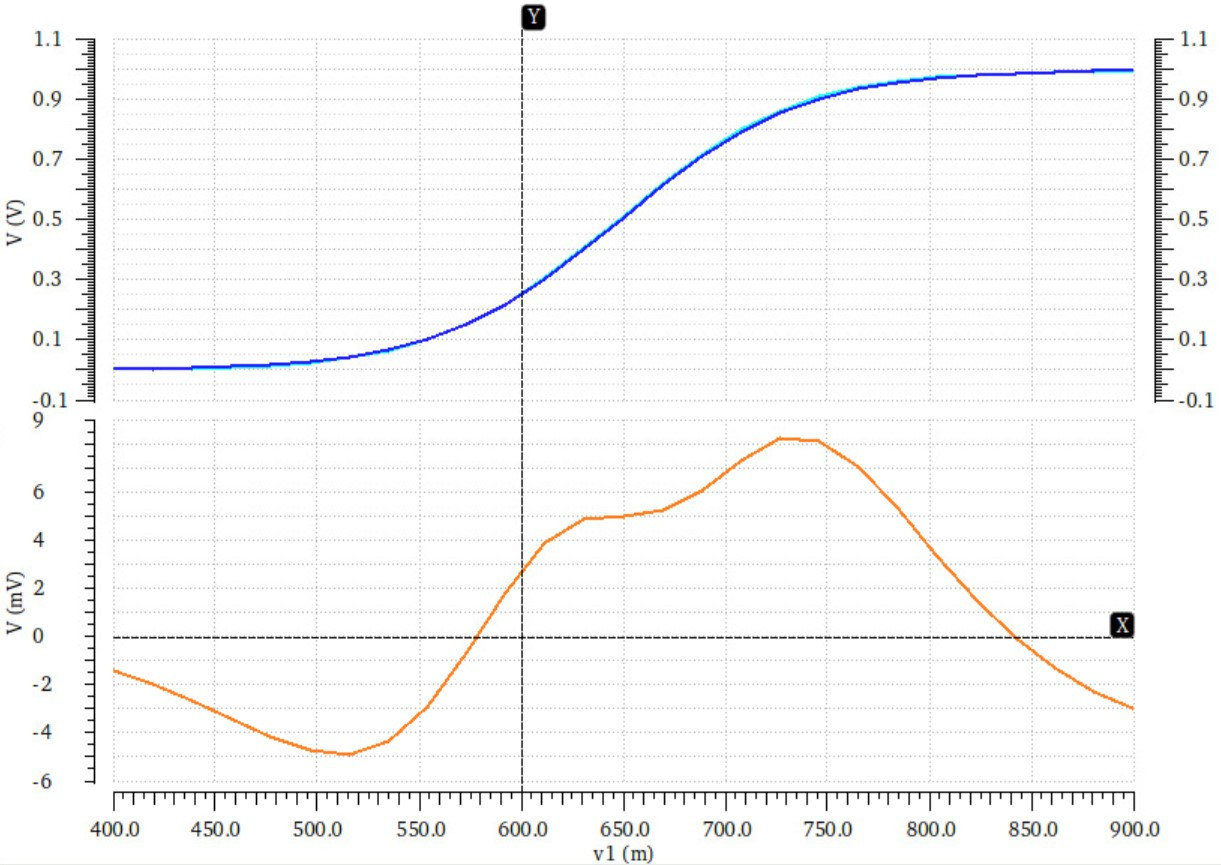}
    \caption{DC Sweep Simulation of the NMOS Processor Topology between 400 mV and 900 mV}
    \label{f15}
\end{figure}

Figure \ref{f15} shows the results of the DC sweep simulation of the NMOS topology. The top plot depicts two curves. In blue, an ideal sigmoid curve was plotted using the ADE-L calculator function. Underneath the curve in light blue, the processor's sigmoid performance with the low-noise load employed. As in the bipolar topology, high accuracy is clear in the large-signal analysis and the lines appear to be right on top of each other. On the bottom plot in orange a sigmoid-error curve is shown, subtracting the ideal sigmoid curve from the measured DC sweep performance of the processor. Two cursors are drawn to show the X-axis of the sigmoid error plot and the 600 mV crossover point as the Y-axis for both plots. The error plot magnitude never exceeds 10 mV, and has an absolute maximum at 8.23 mV, or an inherent error margin of $\pm 0.823\%$.

The inherent error of the NMOS topology is smaller than the bipolar topology by a factor of nearly 1.6 times. This can be due to many factors, but likely most of the accuracy is due to the deployment of the cascode current mirror as the tail current source over the simple current mirror. The simple current mirror has a drastically higher systematic gain error due to a swinging shared emitter/source node, which can correlate to significant swing in the processor accuracy. Additionally, we can see that the sigmoid plot has a horizontal shear compared to the bipolar performance due to the introduction of the subthreshold swing coefficient causing linear stretch in the input-output characteristic.

Transient noise simulations were also conducted in Cadence Virtuoso for the NMOS topology. Using transient data, we were able to capture settling times, rise times, fall times, noise error percentage and signal-to-noise ratios for the topology using both the low-noise loads and PMOS linear-mode loads to compare the performance of each topology. We should note that the definition of settling time for this processor was defined as the time between the start of the output step until the output voltage reaches 99.8 percent of its final value. This is in contrast to the normal settling time definition of zero to 95 percent. This change is made in order to account for ``true settling time" of the processor, as we would not want to sample the output signal until we receive the best possible equilibrium the processor can provide. Due to this fact, in the case of the capacitive low-noise load, the settling time may seem unusually long. 

\begin{figure}
    \centering
    \includegraphics[width=\linewidth]{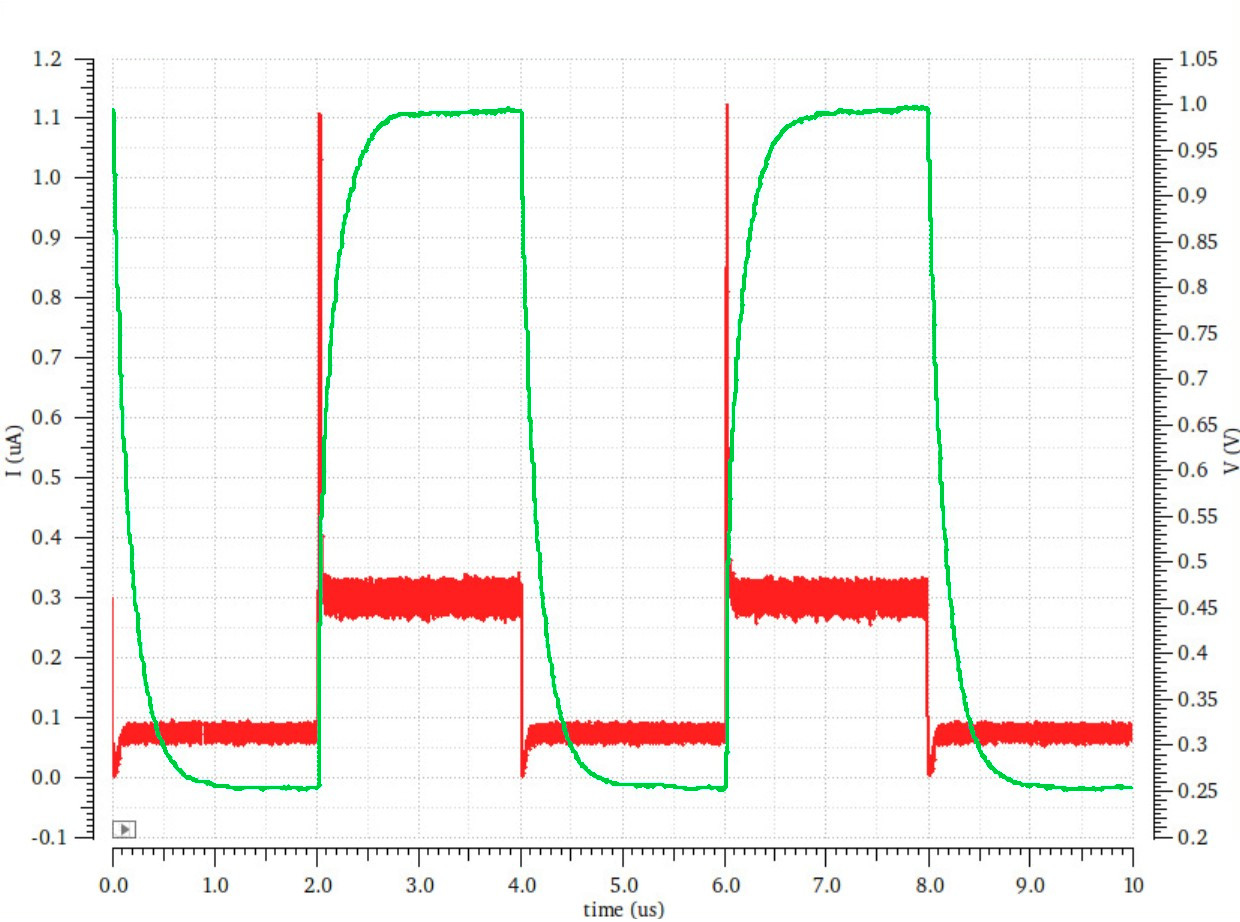}
    \caption{Transient Simulation of the NMOS Processor Topology experiencing a step between a ``well-matched" and ``single-dominant" compute at a frequency of 250 kHz.}
    \label{f16}
\end{figure}

Figure \ref{f16} shows the transient noise simulation of both the output voltage node at the end of the output branch in green and the Softmax branch current on the reference side of the PMOS cascode current mirror in red. An input voltage pulse of 300 mV is driven at the input node at a frequency of 250 kHz and 50 percent duty cycle. This means the input voltage node swings between 600 mV and 900 mV. This simulates a drastic change in the input vector, causing the processor to undergo a transition between a well-matched compute and a single-dominant compute. It can be seen that the inclusion of the noise-dampening load greatly reduces the magnitude of the noise, however at the cost of longer computation times to charge up the 50 fF load capacitance. Designers may opt for a different load topology that garners better computation speed while maintaining large SNR. The table following Figure \ref{f16} shows the side-by-side analysis of the two load cases. Transient simulations of the PMOS linear-mode load are not shown in any figure, but the data in the table was acquired from the simulation test bench.

\vspace{0.1cm}
\begin{center}
    \textbf{Noise Performance of the NMOS Processor}
    \vspace{0.1cm}

    \begin{tabular}{|>{\centering\arraybackslash} m{2cm}|>{\centering\arraybackslash} m{2cm}|>{\centering\arraybackslash} m{2cm}|}
       \hline
       Performance Parameter  &  PMOS Linear Load & Low-Noise Load \\
       \hline
       \hline
       Settling time (99.8\%)  &  72.0 ns & 922.6 ns\\
       \hline
       Rise Time  &  3.2 ns & 360.2 ns\\
       \hline
       Fall Time  &  4.3 ns & 389.4 ns\\
       \hline
       Noise Error (\%)  &  $\pm$16.7\% & $\pm$0.4\%\\
       \hline
       SNR (dB)  &  15.6 dB & 54.0 dB\\
       \hline
    \end{tabular}
\end{center}

\vspace{0.25cm}

There are drawbacks and benefits to using the low-noise load configuration. On the one hand, the settling time to reach an equilibirum solution at the output increases by more than an order of magnitude, which can greatly reduce the maximum operating frequency of the processor. However, the SNR increases by a full 40 decibels. 54.0 dB correlates to approximately 500:1, which means there is roughly a Gaussian variance of 0.2\% in the output voltage. This value is well below the inherent processor error captured in the DC simulations. Meanwhile, 15.6 dB correlates to under 10:1, or a Gaussian variance of 10\%. This level of inconsistency is unacceptable for a high-resolution processor, which greatly necessitates the implementation of a noise-dampening load.

Compared to the bipolar topology, the NMOS circuit consumes far less power. Employing a normal linear-mode PMOS load, the circuit consumes a total of 1.08 $\mu$W. Utilizing the shared cascode PMOS current mirror load, the system consumes 50 percent more power at 1.62 $\mu$W at maximum load.

\section{Experimental Setup}

There are many different factors to take into account when it comes to designing a practical experimental prototype of the analog processor. First, we have to consider the effects of parasitic capacitances that may be inherent in our design that weren't modelled or considered in simulation or ideal circuit analysis. Also, there are other noise sources that weren't accounted for in simulation, such as measurement noise and quantization error that may occur in the data acquisition. Losses due to interconnects are also a factor, as well as overall circuit stability concerns and component mismatch.

To combat mismatch, we recognize that most of the error due to mismatch would occur in the signal transistors. Instead of using individual transistors to construct the signal transistor array, we instead opted for an NPN matched quad transistor array. This ensured that all 4 transistors were manufactured using the same process and in the same die, which can significantly reduce the mismatch between each transistor. For the cascode current mirror we used 4 individual ZTX1048A. To ensure the right amount of current was flowing into the tail current mirror, we employed an ammeter between the shared emitter node and the collector of the current mirror. A potentiometer was used to tune the reference current until the copied mirror current reached the desired tail current within a margin of $\pm$100 $\mu$A. Over time, the resistance of the potentiometer would drift and cause some error in the data acquired from the circuit.

The loads were realized using 20 $\Omega$ resistors. Since the error of the output voltage is directly proportional to the tolerance of the signal-path resistors, we want to choose very low-tolerance resistors. For our design, we purchased resistors with a tolerance of $\pm0.1\%$. Using resistors with tolerance this low ensured that the output error was at most 0.1\%, which is certainly passable for a proof-of-concept experiment.

Next, since we employed a breadboard for the prototype circuit, there are some losses to consider when examining the interconnects. We used conventional 32-gauge tinned-copper jumpers between breadboard ports, which were measured to have between 0.1 $\Omega$ and 0.3 $\Omega$ of intrinsic resistance. Due to this fact, we used as few jumpers as possible in the signal path between the shared emitter node and the load resistors. Losses due to the inclusion of jumpers in other blocks of the circuit would have negligible effect on the output voltage.

On the topic of stability, supply rails were filtered using two 10 $\mu$F capacitors. During initial tests the output voltages all oscillated regardless of input biasing voltage. It took about a week to discover what the cause of this oscillation was, which was due to significant instability caused by the parasitic capacitance between the input, output and shared emitter nodes. Due to the pin configuration of the matched quad, Oscillations could easily travel through the parasitic capacitors from the output nodes to the emitter node, causing a positive feedback loop between the inputs and outputs. This oscillation was mitigated with the introduction of a 300 nF bypass capacitor attached between the emitter node and ground. This allowed high-frequency oscillations on the order of 63 MHz to be filtered out and allow the system to regain stability.

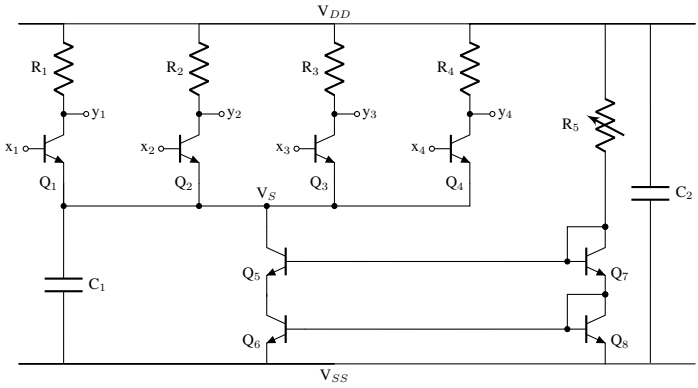
\begin{figure}
    \centering
    \begin{circuitikz}[american, scale=0.6, transform shape]
         \ctikzset{tripoles/mos style/arrows}
         \ctikzset{transistors/arrow pos=end}
         \ctikzset{legacy transistors text}
         \draw (-4,0) node (m1) [npn] {};
         \draw (-1,0) node (m2) [npn] {};
         \draw (2,0) node (m3) [npn] {};
         \draw (5,0) node (m4) [npn] {};
         \draw (0.5,-2.5) node(m5) [npn, xscale=-1] {};
         \draw (0.5,-4) node(m6) [npn, xscale=-1] {};
         \draw (8,-2.5) node(m7) [npn] {};
         \draw (8,-4) node(m8) [npn] {};
         \draw (m1.G) to[short,-o] ++(0,0) node [left] {x$_1$};
         \draw (m2.G) to[short,-o] ++(0,0) node [left] {x$_2$};
         \draw (m3.G) to[short,-o] ++(0,0) node [left] {x$_3$};
         \draw (m4.G) to[short,-o] ++(0,0) node [left] {x$_4$};
         \draw (m1.D) to[short,*-o] ++(0.5,0) node [right] {y$_1$};
         \draw (m2.D) to[short,*-o] ++(0.5,0) node [right] {y$_2$};
         \draw (m3.D) to[short,*-o] ++(0.5,0) node [right] {y$_3$};
         \draw (m4.D) to[short,*-o] ++(0.5,0) node [right] {y$_4$};
         \draw (m1.S) |- ++(3,-0.5) to[short,*-] ++(1.5,0) -- (m5.D) {};
         \draw (m1.S) to[short, -*] ++(0,-0.5) to[C, l=C$_1$] ++(0,-3.5){};
         \draw (9,2.77) to[C, l=C$_2$] ++(0,-7.52){};
         \draw (m2.S) -- ++(0,-0.5);
         \draw (m4.S) |- ++(-3, -0.5) to[short, *-*] ++(-1.5,0) node[above] {V$_S$};
         \draw (m3.S) -- ++(0,-0.5);
         \draw (m1.S) node [left] {Q$_1$};
         \draw (m2.S) node [left] {Q$_2$};
         \draw (m3.S) node [left] {Q$_3$};
         \draw (m4.S) node [left] {Q$_4$};
         \draw (m5.S) node [left, yshift=0.5cm] {Q$_5$};
         \draw (m6.S) node [left, yshift=0.5cm] {Q$_6$};
         \draw (m7.S) node [right, yshift=0.5cm] {Q$_7$};
         \draw (m8.S) node [right, yshift=0.5cm] {Q$_8$};
         \draw (m1.D) to[R, l=R$_1$] ++(0,2) -- ++(-1,0) -- ++(7,0) node [above] {V$_{DD}$};
         \draw (m2.D) to[R, l=R$_2$] ++(0,2);
         \draw (m3.D) to[R, l=R$_3$] ++(0,2);
         \draw (m4.D) to[R, l=R$_4$] ++(0,2) -- ++(5,0) -- ++(-11,0);
         \draw (m7.D) to[short,*-*] ++(0,0) -| (m7.G){};
         \draw (m8.D) to[short,*-*] ++(0,0) -| (m8.G){};
         \draw (m5.G) to[short,-*] (m7.G){};
         \draw (m6.G) to[short,-*] (m8.G){};
         \draw (m7.D) to[vR, l=R$_5$] (8,2.77) {};

         \draw (m6.S) node (vss) [below, xshift=1.5cm] {V$_{SS}$};
         \draw (vss.north) -- ++(-7,0) -- ++(15,0);
     \end{circuitikz}
    \caption{Bipolar Breadboard Prototype Circuit Diagram}
    \label{f17}
\end{figure}

Figure \ref{f17} and the accompanying table shows a circuit diagram of the equivalent breadboard prototype circuit used in testing. A Keysight EDU36311A triple-output programmable DC power supply was used to bias the input nodes and supply voltages. As in the LTSpice and PSpice simulations, $V_{CC}=5$ V and $V_{EE}=0$ V, with the DC bias of the input nodes set to 2.5 V. The DC sweep was measured at the output node using a Keysight DSOX1102G 100 MHz Digital Oscilloscope. The output voltage was measured with respect to ground, so the signals would need to be inverted and properly formatted when extracted from the scope before it could be compared to an ideal sigmoid shape for accuracy measurements.

\begin{tabular}{|>{\centering\arraybackslash} m{2.5cm}|>{\centering\arraybackslash} m{2.5cm}|>{\centering\arraybackslash} m{2.5cm}|}
        \hline
        Designator & Component Value & Details \\ 
        \hline
        \hline
        R$_1$, R$_2$, R$_3$, R$_4$  & 20 $\Omega$ ($\pm0.1\%$), 1/4W & Signal-Path Resistors \\ 
        \hline
        R$_5$ & 1 k$\Omega$ Tunable Potentiometer, 1/2W & Current Mirror Tuning Potentiometer  \\ 
        \hline
        C$_1$ & 300 nF, 25V & Emitter Bypass Capacitor (for stability) \\
        \hline
        C$_2$ & 2$\times$ 10 $\mu$F, 10V & Supply Bypass Capacitors \\
        \hline
        Q$_1$--Q$_4$ & STA412A Quad NPN Array & Matched Transistor Array \\
        \hline
        Q$_5$, Q$_6$, Q$_7$, Q$_8$ & ZTX1048A NPN Transistor & Current Mirror Transistors \\
        \hline
\end{tabular}

\section{Experimental Results}

\begin{figure}
    \centering
    \includegraphics[width=\linewidth]{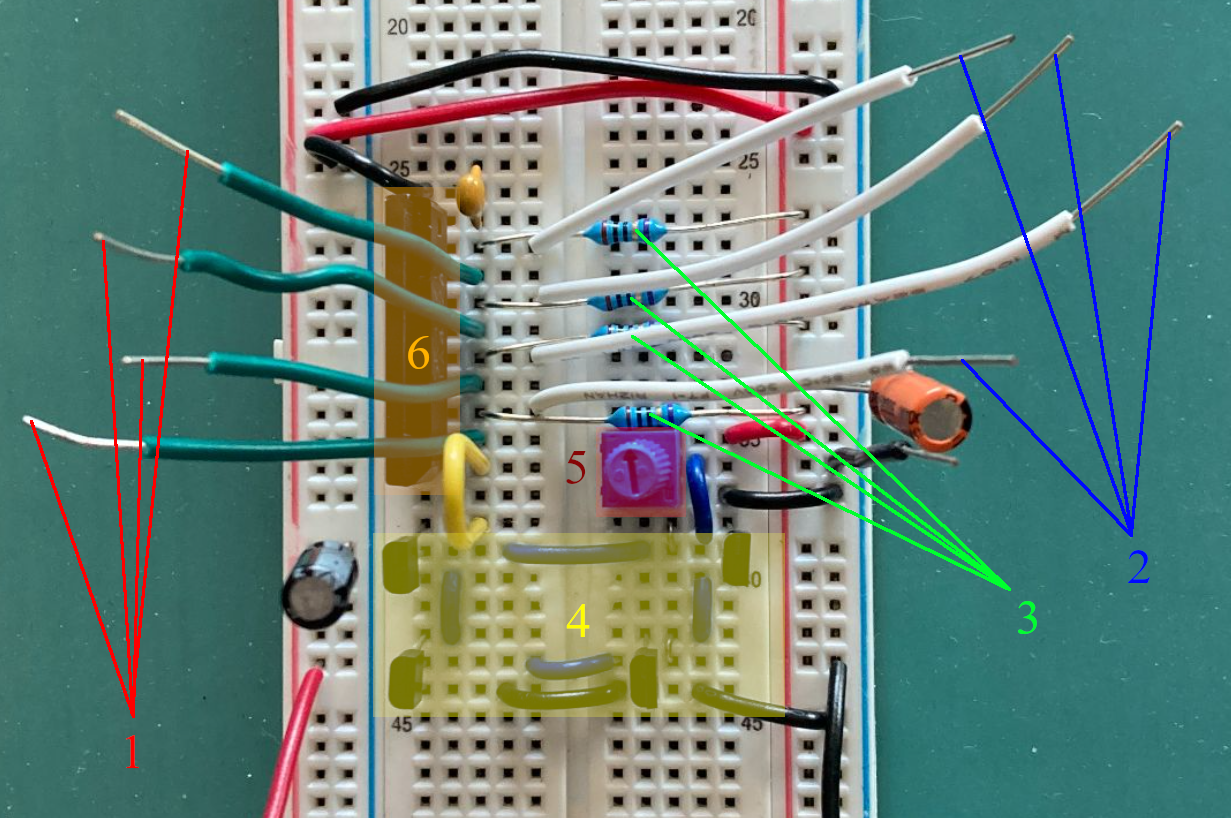}
    \caption{ Image of physical processor breadboard circuit; 1: Input Node Jumper Array, 2: Output Node Jumper Array, 3: Signal-Path Load Resistors, 4: Tail Cascode Current Mirror Block, 5: Reference Current Tuner, 6: NPN Matched Quad.}
    \label{f18}
\end{figure}

Figure \ref{f18} shows a close-up image of the breadboard circuit that was constructed. Labels depict each of the circuit blocks visible in the image. All wires were kept as short as possible to minimize interconnect loss and parasitic capacitance that wasn't inherent in the breadboard. The DC sweep was conducted with a sinusoid with a 1 kHz frequency, and swept roughly between 2.3 and 2.75 volts. Processor accuracy was assessed at 3 different power levels. The output voltage would scale down linearly with power consumption, using tail currents of 50 mA, 25 mA and 5 mA to assess the performance at each power profile.

\begin{figure}
    \centering
    \includegraphics[width=\linewidth]{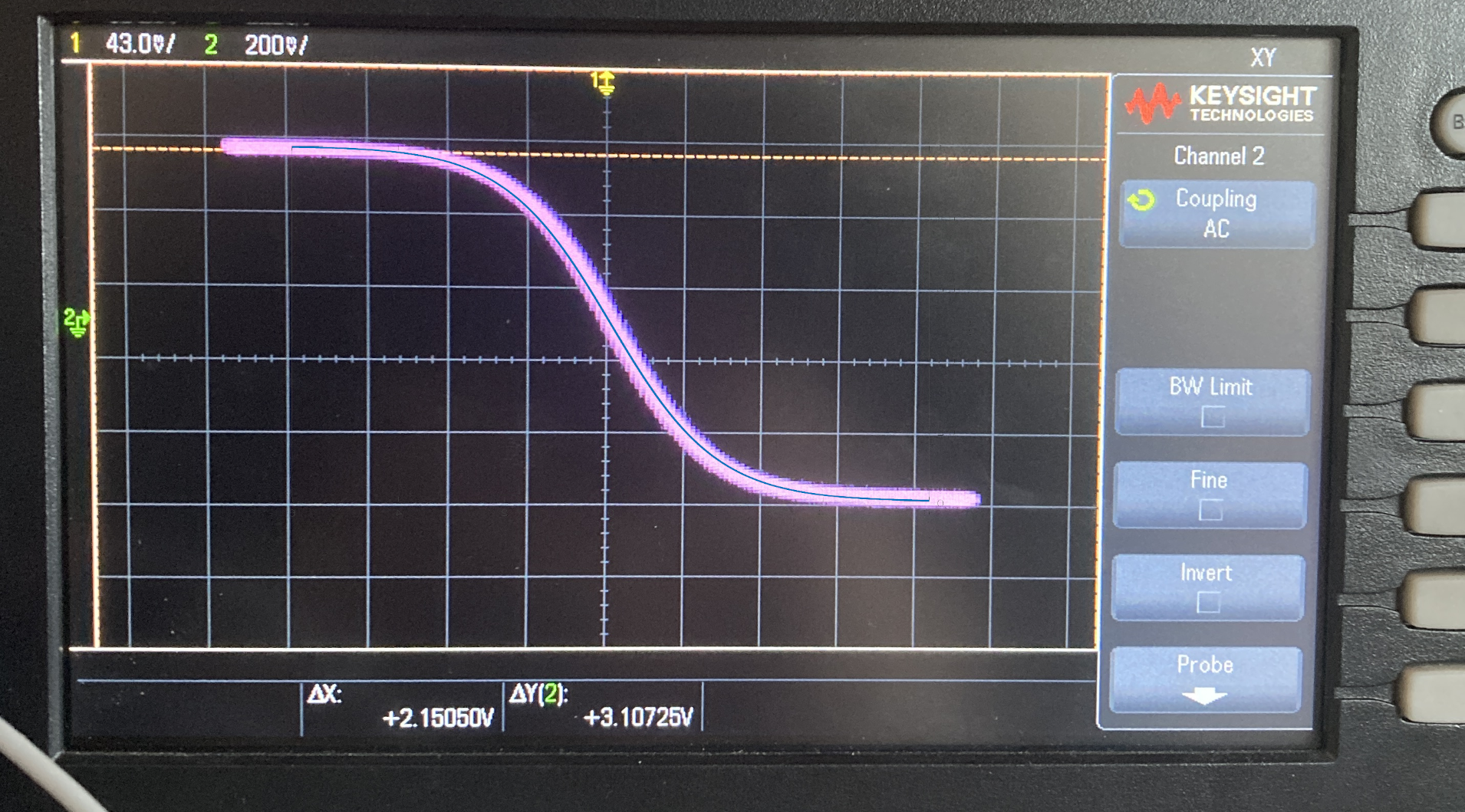}
    \caption{ Sigmoid Curve Superimposed Over the Oscilloscope DC Sweep}
    \label{f19}
\end{figure}

Figure \ref{f19} shows an exemplary sigmoid characteristic curve super-imposed over the output of the oscilloscope. The oscilloscope was used in XY mode to extract the input-versus-output relationship that would depict the sigmoid characteristic. As explained previously, since the output voltage was measured with respect to ground, the sigmoid shape looks reversed compared to simulation data. The data had to be reformatted in Excel in order for the sigmoid shape to face the correct direction, as well as to remove the DC component of the output signal that the oscilloscope extracted.

\begin{figure}
    \centering
    \includegraphics[width=\linewidth]{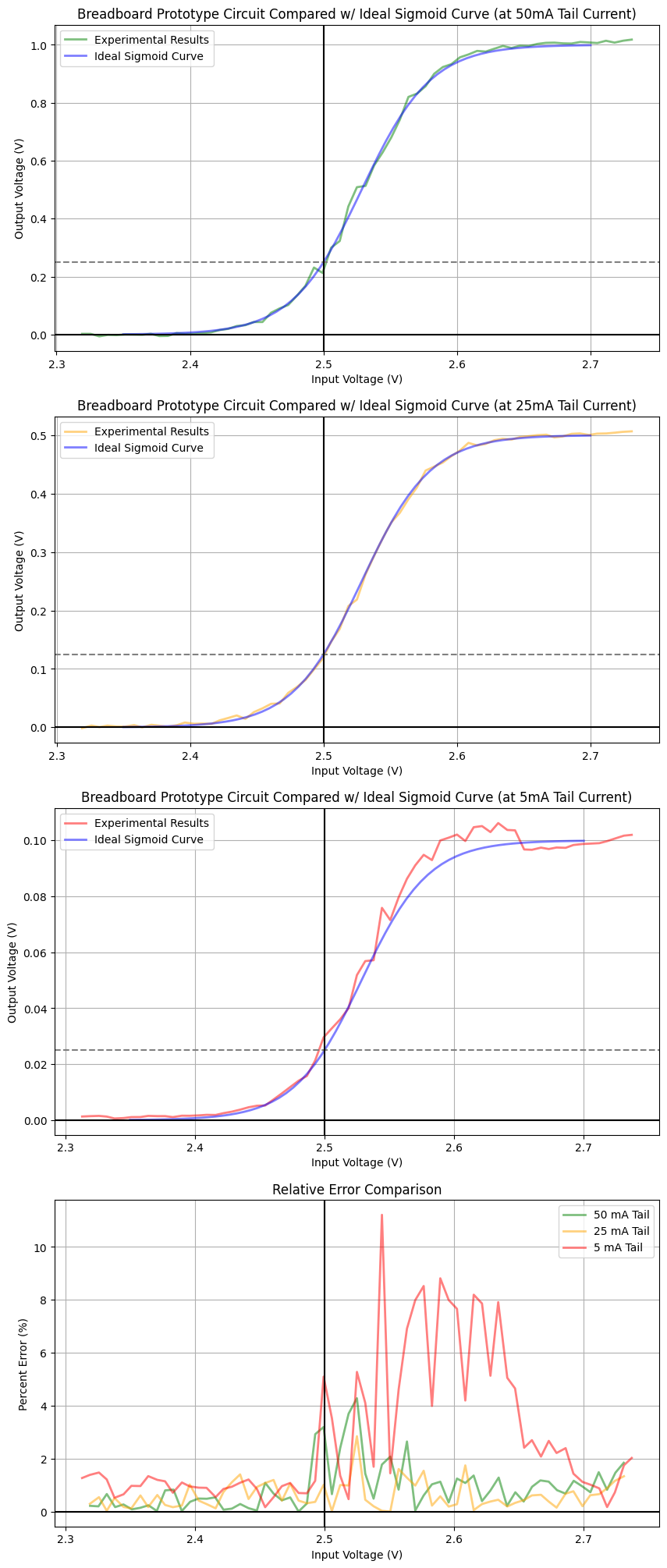}
    \caption{ Experimental Data Performance Analysis Sweeps.}
    \label{f20}
\end{figure}

After the data was properly formatted, the raw oscilloscope data also needed to be shaped. For each x-coordinate in the data, all othe corresponding y-values were averaged to result in a more smooth and straight sigmoid curve. Figure \ref{f20} shows all the data that was extracted for each of the power profiles. Particularly in the case of the lowest power profile (5 mA), there is significant error in the upper-asymptote portion of the extracted data. When viewed on the oscilloscope, there is clearly hysteresis, as on the upwards sweep it overshoots the ideal shape, and on the downward sweep it undershoots the ideal shape. Due to how small the Softmax current is at this power profile, the hysteresis can likely be caused by small variations in $V_{CE,sat}$ between each transistor in the matched quad. Apart from that large contribution to error, it's clear from the relative sigmoid error plot of each power profile that the processor loses no visible accuracy between power states, as we expected from the results of our circuit proof.

Once again, excluding the error due to the hysteresis in the second-half of the sweep for the lowest power profile, we can measure that the error never exceeds more than $\pm4.2\%$. This error would be unacceptable for a final design, but for a proof-of-concept breadboard prototype circuit, this amount of error can be considered to be very low. Considering all of the stacking tolerances, drifting tail current and other nonidealities in the physical design, the processor is remarkably consistent across power profiles and input bias conditions. The accuracy could have likely been vastly improved by instead fabricating a printed circuit board for the processor and using surface-mount components. Constructing the circuit this way would vastly reduce the size of parasitic capacitance, interconnect losses and other parasitic effects. Noise analysis was omitted for the physical design, because it was unlikely that any useful data could be gleaned from the prototype design. The main purpose of the experimental data was to verify the computational accuracy of the topology.

\section{Conclusion and Future Work}

Throughout the previous sections, we have looked over all of the benefits of the analog processor. However, for all the strengths of this design, there are still conceptual and technical caveats that are inherent in the design proposed. 

\subsection{Caveats}

First, the processor topology operates as an open-loop amplifier. Similar to a differential amplifier, the circuit is still based on the negative feedback between the branches that must share the tail current mirror. However, the gain of a differential amplifier is usually further stabilized and made more linear by using a feedback loop between the input and output of the amplifier stage. The circuit topology proposed in this work operates on the opposite assumption, that there is no linear dependence between the input and the output. Therefore the circuit operates in a far less stable state than a comparable amplifier with negative feedback. This means that under certain input biasing conditions, small variations in input biasing can cause large oscillations in output voltage. Thus, the processor operates open-loop and much closer to instability than a system designer may be comfortable with.

Another downside of this topology that was not covered in previous sections is the directly adverse relationship between low power operation and high speed compute. In order for the processor to reach its final equilibrium output values, all nodes must be charged to their final voltage values as well, including the shared source node. The capacitance of all output nodes is relatively low, whereas the capacitance of the shared source node would likely be extremely high. In order for the processor to compute the Softmax function, it must charge or discharge the shared source node in accordance with each new input vector. With the shared source node having a total capacitance of N times the source capacitance of the transistor technology employed, there would be a significant charge time between each compute. So, as the maximum class size of the processor increases, so does the compute time, which creates a direct relationship between class-size and compute time that was previously considered negligible. The only way to reduce the compute time back to a constant value is to increase the current draw of the tail current source in order to charge the shared source node capacitance faster. However, this has the ability to reduce computational accuracy as stated previously.

Lastly, much of today's current high-speed processors operate digitally. As discussed previously, this means that translation between a digital general-purpose processor with this analog processor would require a vast array of digital-to-analog converters. The number of digital signals that would need to be converted would likely be so large that serialization would be required, and each of the analog output voltages would have to be applied to the inputs of the processor sequentially. Once the analog processor block generates the output voltages, the outputs would need to be connected to another large array of analog-to-digital converters, whose outputs would need to be serialized and exported out of the system to be re-integrated in a digital processor. The amount of clock cycles it would take to convert all the digital signals to analog voltages, then back from analog voltages to digital signals would likely take far longer than just computing the Softmax solution using digital logic. Not only will it take a long time to extract the solutions, but the amount of power the ADC and DAC arrays would consume to do the conversions would also dwarf the power consumption of the processing block by a significant degree. Due to this fact, it is important to realize that isolating an analog processor and attempting to integrate it with digital design blocks inevitably results in losses in both power and time to convert between these two domains. It is the opinion of the authors of this paper that this processor will likely find its home in a fully-analog design scheme. AI labs such as IBM \cite{ibm} have already begun redesigning neural network training hardware as fully analog systems by using phase-change memory (PCM) and capacitive MOS arrays to perform matrix multiplication under exceedingly low power conditions. In a full system integration like IBM's, where the system is already completely implemented in an analog design scheme, there would no longer be any power or time losses associated with the use of this Softmax analog processor allowing it to truly shine.

\subsection{Future Work}

There is still much to be done to fully characterize and further improve the processor design in this work. These projects below are examples of a few strong candidates for future works that can build on the data that was presented in this paper. Some of the projects are more hardware-oriented in nature, while others are excellent projects for data scientists and machine learning engineers.

\begin{itemize}
    \item \textbf{Mathematical Derivations}: Skilled electrical engineers can make a short proof that derives comprehensive transfer-function charactersitics of the proposed circuit, including an arbitrary input's bias effect on its own output as well as an arbitrary input's bias effect on all other outputs. This, along with closed-form solutions for input-referred noise equations for the topology will greatly improve the design experience for engineers that wish to implement this circuit topology in any large-scale system designs.
    \item\textbf{Mont\'e Carlo Analysis}: Integrated circuit designers with access to statistical variation transistor models from a foundry with the associated accurate PDKs could run a suite of Mont\'e Carlo simulations to stress the circuit proposed in this work under multiple levels of process variation and mismatch, and compare the accuracy that results. They could also suggest changes to the topology that makes it more robust against process variation, cross-talk and mismatch.
    \item\textbf{Large Class-Size Simulations}: Electrical engineers can run DC sweep and transient simulations on the topology whilst vastly increasing the number of branches being simulated to closely measure how scaling the class-size of the topology affects the compute time and computational accuracy of the circuit. This can help further characterize the practical maximum class size this processor can realistically handle while still maintaining accuracy and low-power operation.
    \item\textbf{Printed Circuit Board Design or Tape-Out}: The best way to truly characterize a circuit is to put designs to work and test the processor as a full-fledged integrated circuit or high-fidelity printed circuit board. Electrical engineers can undertake a tape-out with a trusted foundry and characterize the performance of the integrated circuit that they produce. This can give further credibility to this processor design as well as give high fidelity real world data for the performance of the proposed topology.
    \item\textbf{Improve Accuracy Measurements}: Defining the accuracy of simulations and measurements for an N-dimensional function is a difficult task. Data scientists and firmware design engineers can find more comprehensive ways to define the accuracy of the analog Softmax processor. As an example, they could design a physical and simulation-compatible script and test bench that feeds a large number of example input vectors to the topology and use the resulting output vectors to populate a statistical map of the accuracy of the processor along varying dimensions.
    \item\textbf{Measuring Softmax Inaccuracy Tolerance in Neural Networks}: Software engineers or machine learning engineers can take the relative error from this proposed work and simulate the same level of error in a real Softmax activation function. With this, they can begin to train some rudimentary nerual networks with this ``imperfect" activation function and measure the power consumption, compute time, convergence time and other performance parameters associated with neural networks and compare the results with the same networks that were trained with an ideal Softmax activation function. With this data, they can compare the total power consumed by the ideal activation function process with the theoretical power that the imperfect analog processor would consume.
    \item\textbf{Saturation MOS-based Activation Function}: As a departure from the other suggested projects, this project comes from the discovery of a new activation function that was theorized during the proof of this work. If we instead consider the NMOS topology operating in saturation mode, the function that the processor solves is given by the equation:

    \begin{equation}
        I_{D,i}=\frac{(x_i-G)^2}{\sum_{k=1}^N(x_k-G)^2}\label{eq41}
    \end{equation}

    Where $G$ is given by:

    \begin{equation}
        G=V_S+V_{TH}\label{eq42}
    \end{equation}

    Equation \eqref{eq41} defines a new square-law-based N-dimensional activation function which has not been studied in literature and whose applications are currently unknown. Data scientists and machine learning engineers can take this proposed activation function and characterize its performance compared to other activation functions such as Softmax, ReLU and others in a variety of potential neural network-based tasks.
    
\end{itemize}

\vspace{12pt}

\end{document}